\documentclass[acmsmall,screen]{acmart}

\AtBeginDocument{%
  \providecommand\BibTeX{{%
    \normalfont B\kern-0.5em{\scshape i\kern-0.25em b}\kern-0.8em\TeX}}}

\setcopyright{acmlicensed}
\copyrightyear{2024}
\acmYear{2024}
\acmDOI{XXXXXXX.XXXXXXX}

\acmJournal{CSUR}
\acmVolume{xx}
\acmNumber{x}
\acmArticle{xxx}
\acmMonth{1}

\usepackage{multirow}
\usepackage[caption=false,font=footnotesize,labelfont=rm,textfont=rm]{subfig}
\usepackage{float}
\newtheorem{The}{Theorem}
\newtheorem{Def}{Definition}

\newtheorem{Rem}{Remark}

\newtheorem{Coro}{Corollary}

\begin{document}

\title[Survey of Resilient Coordination]{A Survey of Resilient Coordination for Cyber-Physical Systems Against Malicious Attacks}

\author{Zirui Liao}
\email{by2003110@buaa.edu.cn}
\orcid{0009-0005-7587-7274}
\author{Jian Shi}
\email{shijian@buaa.edu.cn}
\orcid{0000-0002-4475-5195}
\author{Yuwei Zhang}
\email{zhangyuwei@buaa.edu.cn}
\orcid{0000-0001-9384-5173}
\author{Shaoping Wang}
\email{shaopingwang@buaa.edu.cn}
\orcid{0000-0002-8102-3436}
\affiliation{%
  \institution{Beihang University}
  \streetaddress{37 Xueyuan Rd}
  \city{Beijing}
  \country{China}
  \postcode{100191}
}

\author{Zhiyong Sun}
\orcid{0000-0002-4700-1479}
\email{z.sun@tue.nl}
\affiliation{%
  \institution{Eindhoven University of Technology}
  \streetaddress{P.O. Box 513}
  \city{Eindhoven}
  \country{The Netherlands}}

\renewcommand{\shortauthors}{Z.Liao et al.}

\begin{abstract}
  Cyber-physical systems (CPSs) facilitate the integration of physical entities and cyber infrastructures through the utilization of pervasive computational resources and communication units, leading to improved efficiency, automation, and practical viability in both academia and industry. Due to its openness and distributed characteristics, a critical issue prevalent in CPSs is to guarantee resilience in presence of malicious attacks. This paper conducts a comprehensive survey of recent advances on resilient coordination for CPSs. Different from existing survey papers, we focus on the node injection attack and propose a novel taxonomy according to the multi-layered framework of CPS. Furthermore, miscellaneous resilient coordination problems are discussed in this survey. Specifically, some preliminaries and the fundamental problem settings are given at the beginning. Subsequently, based on a multi-layered framework of CPSs, promising results of resilient consensus are classified and reviewed from three perspectives: physical structure, communication mechanism, and network topology. Next, two typical application scenarios, i.e., multi-robot systems and smart grids are exemplified to extend resilient consensus to other coordination tasks. Particularly, we examine resilient containment and resilient distributed optimization problems, both of which demonstrate the applicability of resilient coordination approaches. Finally, potential avenues are highlighted for future research.
\end{abstract}



\received{16 February 2024}

\maketitle

\section{Introduction}
\begin{figure}[t]
    \centering
    \includegraphics[width=0.7\linewidth]{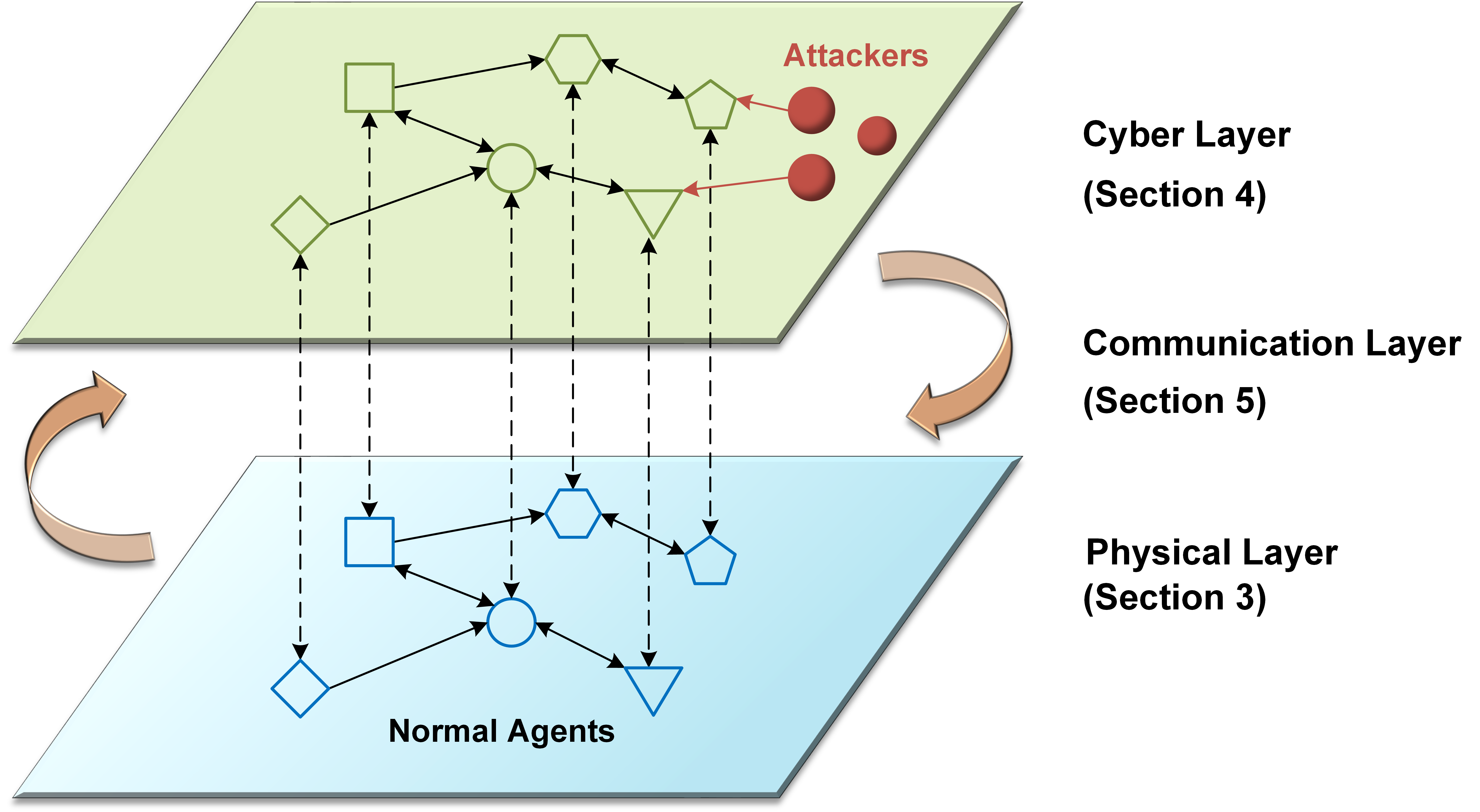}
    \caption{A multilayered framework for CPSs.}
    \label{CPS}
\end{figure}

Combining both physical and computational elements, cyber-physical systems (CPSs) have attracted extensive attention with the development of control theory, communication technique, and signal processing \cite{debie2023swarm,sun2016exponential,li2023survey}. CPSs symbolize the most advanced generation of manufacturing systems with a combination of entities, communication, cyber computing and control abilities \cite{kim2012cyber}. Fig.~\ref{CPS} shows a typical hierarchical framework of the CPS, which consists of the cyber layer, communication layer, and physical layer. The physical layer includes the entities of the CPS and processes the interaction between the CPS and the real world. The cyber layer includes the network topology reflected by the physical layer and computational components of the CPS. The communication layer enables data exchange between different parts of the CPS via various communication mechanisms. As the core of modern industry and manufacturing, CPSs have found widespread and diverse applications in several domains such as mobile robot systems \cite{xu2019networked}, smart grids \cite{xu2018robust}, intelligent transportation systems \cite{wang2023transportation}, and Internet of Things (IoT) \cite{stankovic2014research}. From a macroscopic viewpoint, CPSs are assuming a progressively significant role in the context of the fourth industrial revolution. Globally, various strategic plans have been developed with the introduction of CPSs, e.g., Industry 4.0 (Germany) \cite{angueira2022survey}, Industry Internet (US), Made in China 2025 (China). These initiatives involve CPSs making intelligent decisions in real time and achieving the manufacture of high-quality and customized products with enhanced efficiency on a large scale.

The integration of cyber and physical components in CPSs provides the potential for increased efficiency, automation, and real-time decision making. However, it also brings challenges related to security, privacy, and reliability of various interconnected components and even the entire system. Specifically, (i) the openness of the network medium makes the CPS vulnerable to external malicious attacks; (ii) the distributed characteristic of CPSs makes it difficult to identify and eliminate every potential attack threats; (iii) once the malicious attack is successfully launched, not only the single component may be destroyed but also the entire system will be threatened. These facts have been verified in several security-related events like Stuxnet \cite{masood2019design}, Havex virus \cite{maitra2015offensive}, and RQ-170 attack \cite{hartmann2013vulnerability}. One of their common characteristics is that a single node is infected at first, then misbehavior spreads to the entire system, leading to system abnormality or paralysis.

In recent years, an increasing number of research and review articles have investigated the characterization and influence of malicious attacks. Node injection attack is a typical malicious attack which has been widely studied in \cite{ishii2022overview,yan2020resilient,fu2021resilient}. The authors in \cite{ishii2022overview} classified the node injection attack into a false data injection (FDI) process and reviewed fundamental results for achieving resilient consensus. In \cite{yan2020resilient}, the node injection attack was characterized as an adversarial environment and resilient containment problems for first-order and second-order multi-agent systems (MASs) were studied. In \cite{fu2021resilient}, the node injection attack was modeled as a kind of deception attack occurred on agents and the resilient distributed optimization problem was investigated. In addition, the survey paper \cite{ghasempouri2023survey} provides a comprehensive overview of existing architectural attacks, which leverage hardware vulnerabilities to launch software attacks. For aircraft security reasons, the authors in \cite{habler2023assessing} summarize common malicious attacks on aviation systems and components and examine their impact on these components and systems.

The aforementioned security concerns motivate us to design reliable and secure controllers and algorithms, thereby achieving a desired global objective when the network is subject to malicious attacks. In response to these severe malicious attacks that have occurred around the world, the notion of \textit{resilient coordination} was developed, which aims to guarantee that the CPS maintains its essential functions and services, even when faced with adverse conditions or challenges. The research on resilient algorithms has also achieved rapid development. Generally, there are two categories of methods for addressing resilient coordination problems. One category is based on the idea of fault identification and isolation, i.e., adversary agents in the network are firstly identified, and then isolated \cite{tang2019linear,jia2020partial}. For distributed CPSs, however, identifying and isolating adversary agents are challenging since these processes require agents to handle massive information. It is also impractical to identify all malicious attacks in large-scale distributed CPSs. 

The second category comprises a set of algorithms known as  \textit{Mean-Subsequence-Reduced} (MSR), which has garnered significant interest over the past decade owing to its fault-tolerant and light-weight characteristics. In \cite{leblanc2013resilient}, the MSR algorithm was adopted to achieve resilient consensus against malicious and Byzantine attacks, with each normal agent filtering the suspicious values received from its in-neighbors. Variants of this algorithm in different scenarios include the weighted MSR (W-MSR) \cite{leblanc2013resilient}, sliding-window MSR (SW-MSR) \cite{saldana2017resilient}, event-based MSR (E-MSR) \cite{wang2019resilient}, double-integrator position-based MSR (DP-MSR) \cite{dibaji2017resilient}, and quantized-weighted MSR (QW-MSR) \cite{dibaji2017resilient2} algorithms. Considering its numerous advantages, this survey will focus on reviewing and discussing the promising results of resilient coordination using MSR-type algorithms.

In recent years, there emerge several survey papers \cite{ishii2022overview,he2021secure,aslam2022overview,pirani2022survey,wang2023resilient,zhang2022reaching} that focus on establishing secure and resilient CPSs. The authors in \cite{ishii2022overview} provide a comprehensive examination of resilient consensus in adversarial environments, where different attack models and corresponding resilient strategies are discussed from a cybernetic perspective. In \cite{he2021secure}, the authors present a multi-layered framework for the MAS and classify malicious attacks according to different configuration layers. Subsequently, several distributed secure controllers are surveyed to defend against malicious attacks. The paper \cite{aslam2022overview} presents several typical malicious attacks, including the deception attack, Denial of Service (DoS), and replay attack. Subsequently, the detection and resilient strategies against these attacks are further developed. The work \cite{pirani2022survey} analyzes the resilience of networked control systems (NCSs) based on graph-theoretic methodologies. A comparative survey is conducted in \cite{wang2023resilient}, which reviews state-of-the-art findings of cyber threats on MASs and resilient coordination strategies. A complete survey on Byzantine fault-tolerant consensus algorithms is given in \cite{zhang2022reaching}.

The in-depth security analysis of CPSs depend on a comprehensive understanding to system framework. However, most existing reviews adopt the attack type as the classification criteria, while neglecting some situations that CPSs may be confronted with in practice. These situations include: (i) various physical structures and models due to the heterogeneity of the CPS; (ii) restricted interaction due to limited communication and computation resources; and (iii) time-varying network topologies due to uncertain environments. The locations where these situations occur correspond to the physical layer, network layer, and cyber layer in the multi-layered framework of the CPS, respectively. In addition, previous surveys mainly focused on multi-agent consensus control under adversarial environments, leading to the omission of other application-oriented resilient coordination tasks that have been hot topics in recent years.

\begin{table*}[!t]
\caption{An overview of recent surveys on resilient coordination.}
\label{Tab1}
\centering
\resizebox{1\linewidth}{!}{
\begin{tabular}{ll|l|l|l}
\hline
\multicolumn{1}{c|}{Year} & Survey   & Classification Criterion            & Concerned Problems                  & Platform     \\ \hline
\multicolumn{1}{l|}{2022} & Ishii \textit{et al.} \cite{ishii2022overview} & Attack model & Resilient consensus  & MASs \\ \hline
\multicolumn{1}{l|}{2022} & He \textit{et al.} \cite{he2021secure} & Attack model & Resilient consensus & MASs          \\ \hline
\multicolumn{1}{l|}{2022} & Aslam \textit{et al.} \cite{aslam2022overview} & Attack model & Resilient consensus & MASs          \\ \hline
\multicolumn{1}{l|}{2023} & Pirani \textit{et al.} \cite{pirani2022survey} & Graph-theoretic approaches & Resilient coordination & NCSs          \\ \hline
\multicolumn{1}{l|}{2023} & Wang \textit{et al.} \cite{wang2023resilient} & Attack model & Resilient consensus & MASs          \\ \hline
\multicolumn{1}{l|}{2024} & Zhang \textit{et al.} \cite{zhang2022reaching} & Algorithm type & Resilient consensus & Blockchains          \\ \hline
\multicolumn{2}{c|}{This survey}     & A multi-layered framework&  Resilient coordination & CPSs \\ \hline
\end{tabular}
}
\end{table*}

\begin{figure*}[!t]
    \centering
    \includegraphics[width=0.85\linewidth]{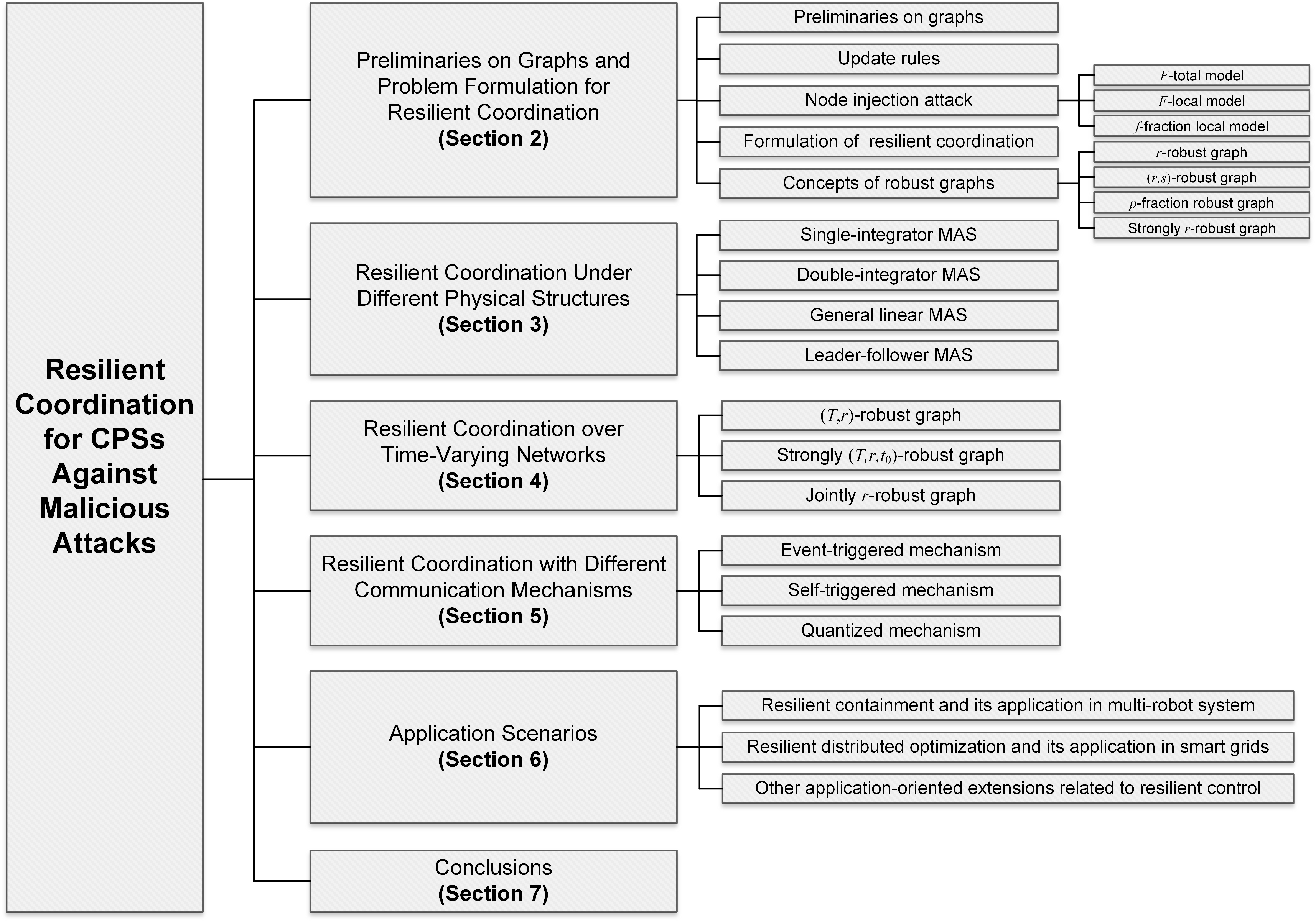}
    \caption{Main structure of this survey.}
    \label{ReviewStructure}
\end{figure*}

A comprehensive comparison between the existing survey papers and this review is summarized in TABLE~\ref{Tab1}. To the best of the authors' knowledge, existing research lacks a comprehensive literature analysis about the advancements for CPSs considering the underlying framework of the system and various resilient coordination tasks beyond consensus. Thus, it is essential to conduct a thorough review of the latest developments in resilient coordination for CPSs and provide a detailed statement of the graph conditions and limitations associated with these results. Compared with the existing review papers, this survey proposes a comprehensive and in-depth taxonomy that prioritizes the framework of CPS, and emphasizes the resilient coordination problems for MASs against the node injection attack.\footnote{Essentially, while an MAS represents a physical structure for interaction and decision-making among autonomous agents, a CPS encompasses the integration of computational elements with physical entities. Thus, MAS can be employed within CPS for better coordination, control, and optimization.} Some other application-oriented extensions related to resilient coordination are further reviewed with the corresponding literature for reference. To start with, some preliminaries on graphs and fundamental problem setting for resilient coordination are stated. Subsequently, promising results of resilient coordination are classified and reviewed from three perspectives: physical structure, communication mechanism, and network topology. Furthermore, two typical application scenarios are exemplified to extend resilient consensus to other coordination tasks and demonstrate the applicability of the resilient coordination approaches. Finally, some potential avenues are highlighted for future research. The main structure of this survey is depicted in Fig.~\ref{ReviewStructure}.

\section{Preliminaries and Problem Formulation for Resilient Coordination}
\label{sec2}
\subsection{Preliminaries on Graphs}
Consider a graph $\mathcal{G}=(\mathcal{V}, \mathcal{E})$ consisting of $n$ nodes. The node set is given as $\mathcal{V}$, with $|\mathcal{V}|$ being its cardinality. The edge set is denoted as $\mathcal{E} \subseteq \mathcal{V} \times \mathcal{V}$. The edge $(j, i) \in \mathcal{E}$ means that node $i$ has access to the information of node $j$. Denote the sets of in-neighbors and out-neighbors for agent $i$ as $\mathcal{V}_{i}^{+}=\{j \in \mathcal{V}|(j, i) \in \mathcal{E}\}$ and $\mathcal{V}_{i}^{-}=\{j \in \mathcal{V}|(i, j) \in \mathcal{E} \}$, respectively. The adjacency matrix $\mathcal{A}_{\mathcal{G}}=\left[a_{i j}\right] \in \mathbb{R}^{n \times n}$ associated with $\mathcal{G}$ is defined as $a_{i j} \in (0, 1)$ if $(j, i) \in \mathcal{E}$ and otherwise $a_{i j}=0$, where $a_{ij}$ is the weight of the edge $(j, i)$ and $i,j \in \mathcal{V}$.

\subsection{Update Rules Under Different System Structures} 

Resilient consensus typically seeks to defend adversarial attacks and attain a global objective for the states or outputs of MASs by designing secure controllers, which are executed merely through the local information interaction. According to different system structures, resilient consensus problems are divided into leaderless resilient consensus problems \cite{leblanc2013resilient, peng2023resilient, wen2023joint, yuan2023event} and leader-following resilient consensus problems \cite{usevitch2018resilient, usevitch2019resilient, rezaee2021resiliency, zegers2019event, yan2020resilient}. Correspondingly, the update rules for agents in leaderless and leader-follower structures are presented below.

\subsubsection{Leaderless structure} 
Leaderless resilient coordination indicates that all agents behave in a cooperative and distributed manner to attain a global objective. In a leaderless structure, each agent has the equal status and there is no designated leader agent in the system.

Assume that each agent possesses an initial state, which can be interpreted as a measurement, optimization variable, etc.\footnote{Although the state values are usually formatted in one-dimensional space, most results presented in this survey can be extended to multi-dimensional space through Kronecker product.} Agents engage in synchronous information interaction by sending their state values to out-neighbors and receiving state values from their in-neighbors. The state update for each agent relies on the information received from its in-neighbors and follows a specific protocol. Note that the update protocol is a flexible function that can vary for each agent, based on its specific cooperative task. In general, update protocols are pre-designed in such a way that the agents in the network perform a desired computation.

\subsubsection{Leader-follower structure} 
The leader–follower resilient coordination is based on a leader-follower structure that receives lots of attention in recent years \cite{franze2020resilient}. In numerous real-world scenarios, it is usually advantageous to possess a leader, either single or multiple, either static or dynamic, to assign a global objective or collective behavior for the entire group of agents. Generally, a leader-follower MAS contains two types of agents: 1)~leader agents (leaders); and 2)~follower agents (followers), with the former propagating desired outputs and the latter following these reference signals to complete cooperative control tasks. The state update for followers also relies on the information received from their in-neighbors, while the state update for leaders does not require information interaction with other nodes. Instead, their update protocols are prescribed by designers based on specific objectives. 

Coordination of CPSs without malicious attacks has been extensively studied under leaderless \cite{ren2005consensus, ren2009distributed, sun2018cooperative, zhao2017general} and leader-follower \cite{arcak2007passivity, abdessameud2016leader, wang2017distributed, zhang2022distributed, zhang2023self, zhang20233d} structures. Nevertheless, due to the distributed characteristic of the system and openness of the communication network, CPSs are susceptible to malicious attacks. For example, the normal agents may be manipulated by attackers and become misbehaving agents, whose states deviate from the prescribed update rules. These misbehaving agents apply some other abnormal rules for state update and disseminate malicious information to their out-neighbors. Such misbehavior may diminish the effectiveness of control strategies and disrupt system security. Therefore, it is crucial to design secure protocols so that the effect of misbehaving agents can be reduced or even eliminated.

\subsection{Node Injection Attack Model} 
\label{subsec2.3}
The aforementioned security concerns prompt us to explore the nature and specific model of malicious attacks. In this review, we mainly focus on node injection attack. The attackers leverage their knowledge about the underlying network to target the vulnerable components of a CPS, thus they aim to maximize their influence and minimize their visibility. Successful attack gives full control over the target system to the adversary \cite{andrea2015internet}. Thus, this kind of attack not only compromises individual nodes, but also threatens the security of the entire system.

\begin{figure*}[!t]
    \centering
    \includegraphics[width=0.85\linewidth]{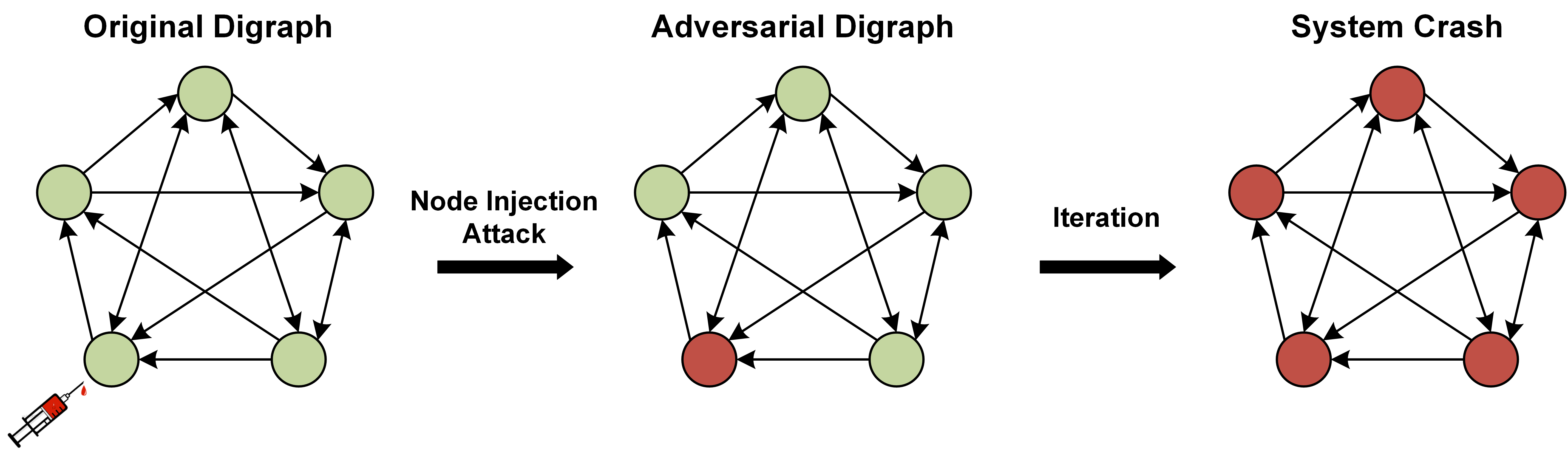}
    \caption{The evolution of malicious attacks injected by a single node.}
    \label{NodeInjectionAttack}
\end{figure*}

The attack injection process is illustrated in Fig.~\ref{NodeInjectionAttack}. Firstly, a vulnerable node in the original graph is injected with malicious information, and the healthy network becomes the adversarial network. Subsequently, the injected node is manipulated by the attacker and acts as an adversary agent. Through iterations of time, it will transmit malicious information to all of its out-neighbors and infect them, thereby threatening the entire system. As shown in Fig.~\ref{NodeInjectionAttack}, even the single node injection attack has the ability of causing the system to crash eventually.

In the context of node injection attack, agents in the CPS are divided into normal agents and adversary agents, with the former collaborating with in-neighbors to achieve resilient coordination tasks and the latter transmitting wrong information to out-neighbors to interrupt the normal system update. Adversary agents are further subdivided into malicious and Byzantine agents. The detailed definitions of normal, malicious, and Byzantine agents are presented as follows.

\begin{Def}[Normal agent]
An agent is normal if it updates the state according to the predetermined rule and sends its true state value to all of its out-neighbors at each time step.
\end{Def}

\begin{Def}[Malicious agent]
An agent is malicious if it sends its true state value to all of its out-neighbors at each time step, but does not adhere to the predetermined update rule at some time steps.
\end{Def}

\begin{Def}[Byzantine agent]
An agent is Byzantine if it does not send the same value to all of its out-neighbors at some time steps, or if it does not adhere to the predetermined update rule at some time steps.
\end{Def}

The sets of normal, malicious, and Byzantine agents are denoted as $\mathcal{N}$, $\mathcal{M}$, and $\mathcal{B}$, respectively. According to the aforementioned definitions, we know that the main differences among these three kinds of agents are whether they adhere to predetermined update rules and whether they broadcast their true state values. Furthermore, all malicious agents are Byzantine, but not necessarily vice versa. Note that malicious agents are generally appropriate for modeling broadcasting networks (e.g., wireless sensor network \cite{kikuya2017fault} and robotic networks \cite{guerrero2018design}), while Byzantine agents are more common in peer-to-peer networks \cite{kim2010counteracting,chang2022byzantine}.

The definitions of different agents also induce the set of adversary agents as $\mathcal{A}=\mathcal{M} \cup \mathcal{B}$. Generally, it is assumed that $\mathcal{A} \cap \mathcal{N}=\varnothing$ and $\mathcal{A} \cup \mathcal{N}=\mathcal{V}$. To better describe the influence of malicious attacks, the paper \cite{leblanc2013resilient} has defined the following three attack models.

\begin{Def}[$F$-total model]
A multi-agent network is said to be an $F$-total model if the whole network possesses at most $F$ adversary agents, i.e., $\left|\mathcal{A}\right| \leq F$.
\end{Def}

\begin{Def}[$F$-local model]
A multi-agent network is said to be an $F$-local model if the in-neighbor set of each agent $i$ contains at most $F$ adversary agents at each time step $t$, i.e., $\left|\mathcal{V}_i^+[t] \cap \mathcal{A}\right| \leq F, \ \forall i \in \mathcal{V}$.
\end{Def}

\begin{Def}[$f$-fraction local model]
A multi-agent network is said to be an $f$-fraction local model if the in-neighbor set of each agent $i$ contains at most a fraction $f$ of adversary agents at each time step $t$, i.e., $\left|\mathcal{V}_i^+[t] \cap \mathcal{A}\right| \leq f\left|\mathcal{V}_i^+[t]\right|, \forall i \in \mathcal{V}$, $f \in [0,1]$.
\end{Def}

\begin{figure*}[!t]
    \centering
    \includegraphics[width=0.95\linewidth]{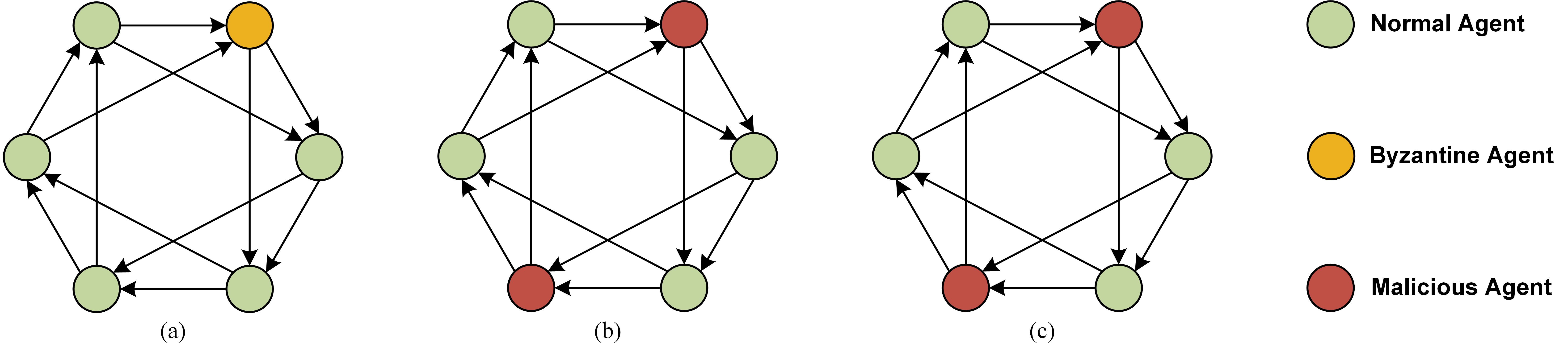}
    \caption{Illustrations of different threat models: (a) $1$-total Byzantine model, (b) $1$-local malicious model, and (c) $0.5$-fraction local malicious model.}
    \label{attack}
\end{figure*}

Illustrations of the $1$-total malicious model, $1$-local Byzantine model, and $0.5$-fraction local malicious model are presented in Figs.~\ref{attack}(a), \ref{attack}(b), and \ref{attack}(c), respectively. By comparing Fig.~\ref{attack}(b) and Fig.~\ref{attack}(c), we find that they share the same number of adversary agents. This fact indicates that different threat models are sometimes interchangeable.

Conventional distributed control methodologies implicitly assume that all agents operate reliably and work collaboratively to attain a global objective. However, the design of distributed algorithms rely heavily on communication infrastructures, which create numerous vulnerabilities for cyber attacks. In such attacks, external adversaries may manipulate the transmitted information. Moreover, it is possible for an adversary agent to intentionally disseminate self-generated data to out-neighbors with malicious purposes. It is evident that such misbehaviors undermine the performance of distributed protocols by impeding normal agents from reaching the expected state value or manipulating the final state value to be false. More seriously, a single adversary agent may compel all agents to reach arbitrary state values by merely keeping this value constant, thereby failing to achieve the preset global objective. Thus, it is critical to study resilient coordination for CPSs by designing reliable and secure algorithms to achieve the desired control objective in the presence of node injection attack.

\subsection{Formulation of the Resilient Consensus Problem}
As a basic problem in the field of control, distributed consensus has gained extensive attention in recent decades due to its broad range of applications \cite{olfati2007consensus}. In the context of distributed consensus, each agent in the network interacts and shares the information only with its local in-neighbors, and the whole system should reach an agreement at a steady state \cite{ren2007information}. In principle, all of the studies on consensus without malicious attacks can be extended with the consideration of security risks, and corresponding resilient distributed algorithms should be also considered. Motivated by the this principle, the notion of \textit{resilient distributed algorithm} is presented below \cite{pirani2022survey}.
\begin{Def}[\textit{Resilient distributed algorithm}]
\label{RDA}
   Suppose that a multi-agent network satisfies a certain attack model. A distributed algorithm is said to be resilient if each normal agent attains its desired objective by implementing the algorithm, regardless of the misbehavior of adversary agents.
\end{Def}

Following the idea of Definition~\ref{RDA}, miscellaneous resilient problems and corresponding resilient distributed algorithms can be considered according to different system structures, network topologies, and communication mechanisms. We start with the formulation of the fundamental resilient consensus problem, which is characterized as an asymptotic property.

\begin{Def}[\textit{Resilient consensus}]
\label{Problemerc}
    Suppose that the network satisfies a certain attack model ($F$-total, $F$-local, or $f$-fraction local). For any choice of initial values, determine graph conditions and design controllers such that the following resilience and consensus conditions hold.
\begin{itemize}
    \item Resilience condition: For each normal agent $i \in \mathcal{N}$ and for all time steps $t \in \mathbb{Z}_{\geq 0}$, it holds $x_i[t] \in \mathcal{S}$, where $\mathcal{S}=\left[\min _{i \in \mathcal{N}} x_i[0], \max _{i \in \mathcal{N}} x_i\left[0\right]\right]$.
    \item Consensus condition: For each pair of normal agents $i, j \in \mathcal{N}$ and for time step $t \in \mathbb{Z}_{\geq 0}$, it holds $\lim _{t \rightarrow \infty} \left|x_i[t]-x_j[t] \right| = 0$.
\end{itemize}
\end{Def}

As the extension of non-resilient cases \cite{ren2007information,mesbahi2010graph}, the formulation of (exact) resilient consensus problem consists of two essential conditions. The resilience condition is to keep the state values of the normal agents always within a safe range. The consensus condition ensures that the normal agents asymptotically achieve exact consensus in presence of node injection attacks.

In some cases where exact consensus is unrealizable, approximate consensus will be considered instead \cite{wang2019resilient,yan2020resilientevent,yuan2023event}. The corresponding consensus condition is modified as below.
\begin{Def}[\textit{Approximate resilient consensus}]
\label{Problemarc}
    Suppose that the network satisfies a certain attack model. For any choice of initial values, determine graph conditions and design controllers such that the following resilience and consensus conditions hold.
\begin{itemize}
    \item Resilience condition: The same as the resilience condition formulated in Definition~\ref{Problemerc}.
    \item Consensus condition: For each normal agent $i \in \mathcal{N}$, it holds $\lim _{t \rightarrow \infty} \left|x_i[t]-x_j[t] \right| \leq c, \ \forall i, j \in \mathcal{N}$, where $c$ is a positive error range to be achieved by normal agents.
\end{itemize}
\end{Def}

 Generally, the resilience condition is implicitly assumed to be satisfied, with only the consensus condition being explicitly imposed. Both Definition~\ref{Problemerc} and Definition~\ref{Problemarc} establish on the assumption that the network is subject a specific attack model. For unbounded attacks (any number of adversaries), the paper \cite{abbas2014resilient}  proved that a connected dominating set is required to ensure resilient consensus, which a relatively conservative condition. In addition to the aforementioned two problems, there are some variants of the resilient consensus problem, which will be discussed in detail later.



\subsection{Robust Network Topologies}

Some essential notions with respect to (w.r.t.) sets and
graphs are presented to characterize the resilient properties for the CPS, i.e., reachability and robustness. Intuitively, these properties can be regarded as derivatives of the three attack models defined in Section~\ref{subsec2.3} and are presented below.
\begin{Def}[\textit{$r$-reachable set}]
 Consider a graph $\mathcal{G}=(\mathcal{V}, \mathcal{E})$ and a nonempty subset $\mathcal{S} \subseteq \mathcal{V}$. $\mathcal{S}$ is $r$-reachable if $\exists \, i \in \mathcal{S}$ such that $\left|\mathcal{V}_i^+ \right \backslash \mathcal{S} \mid \geq r$, where $r \in \mathbb{Z}_{>0}$.
\end{Def}

\begin{Def}[\textit{$(r,s)$-reachable set}]
Consider a graph $\mathcal{G}=(\mathcal{V}, \mathcal{E})$ and a nonempty subset $\mathcal{S} \subseteq \mathcal{V}$. $\mathcal{S}$ is $(r,s)$-reachable if given $\mathcal{Y}_s^r=\left\{i \in \mathcal{S}:\left|\mathcal{V}_i^+\right \backslash \mathcal{S} \mid \geq r\right\}$, then $\left|\mathcal{Y}_s^r\right| \geq s$, where $r, s \in \mathbb{Z}_{>0}$.
\label{def3}
\end{Def}

\begin{Def}[\textit{$p$-fraction reachable set}]
Consider a graph $\mathcal{G}=(\mathcal{V}, \mathcal{E})$ and a nonempty subset $\mathcal{S} \subseteq \mathcal{V}$. $\mathcal{S}$ is $p$-fraction reachable if $\exists \, i \in \mathcal{S}$ such that $\left|\mathcal{V}_i^+ \right \backslash \mathcal{S} \mid \geq p \left|\mathcal{V}_i^+ \right|$, where $p \in [0,1]$.
\end{Def}


The above notions can be extended to graphs and the following definitions are derived.

\begin{Def}[\textit{$r$-robust graph}]
Consider a graph $\mathcal{G}=(\mathcal{V}, \mathcal{E})$. $\mathcal{G}$ is $r$-robust if for each pair of nonempty, disjoint subsets $\mathcal{S}_1,\mathcal{S}_2 \subseteq \mathcal{V
}$, at least one of them is $r$-reachable, where $r \in \mathbb{Z}_{>0}$.
\label{defrrobust}
\end{Def}

\begin{Def}[\textit{$(r,s)$-robust graph}]
Consider a graph $\mathcal{G}=(\mathcal{V}, \mathcal{E})$ with $n$ ($n\geq2$) nodes. $\mathcal{G}$ is $(r, s)$-robust if at least one of the conditions given below is satisfied specific to each pair of nonempty, disjoint subsets $\mathcal{S}_1,\mathcal{S}_2 \subseteq \mathcal{V
}:$
\begin{equation*}
1) \left|\mathcal{Y}_{\mathcal{S}_1}^r\right|=\left|\mathcal{S}_1\right|, \quad
2) \left|\mathcal{Y}_{\mathcal{S}_2}^r\right|=\left|\mathcal{S}_2\right|, \quad
3) \left|\mathcal{Y}_{S_1}^r\right|+\left|\mathcal{Y}_{S_2}^r\right| \geq s,
\end{equation*}
where $r \in \mathbb{Z}_{>0}$, $1 \leq s \leq n $, $\mathcal{Y}_{S_p}^r$ $(p=1,2)$ is the node set in $\mathcal{S}_p$ with at least $r$ in-neighbors outside of $\mathcal{S}_p$, which is expressed as $\mathcal{Y}_{\mathcal{S}_p}^r=\left\{i \in \mathcal{S}_p:\left|\mathcal{V}_i^{+}\right \backslash \mathcal{S}_p \mid \geq r\right\}$.
\label{def4}
\end{Def}

\begin{Def}[\textit{$p$-fraction robust graph}]
Consider a graph $\mathcal{G}=(\mathcal{V}, \mathcal{E})$. $\mathcal{G}$ is $p$-fraction robust if for each pair of nonempty, disjoint subsets $\mathcal{S}_1,\mathcal{S}_2 \subseteq \mathcal{V}$, at least one of them is $p$-fraction reachable, where $p \in [0,1]$.
\label{defpfrac}
\end{Def}

Under a leader-follower structure, a stronger graph robustness w.r.t.~a subset $\mathcal{S}_1 \subseteq \mathcal{V}$ (where $\mathcal{S}_1$ is usually assigned to be the set of leaders $\mathcal{L}$) is required for achieving resilient coordination. Therefore, Definition~\ref{defrrobust} is extended as follows \cite{mitra2016secure}.

\begin{Def}[\textit{strongly $r$-robust graph}]
Consider a time-invariant graph $\mathcal{G}=(\mathcal{V}, \mathcal{E})$ and a nonempty subset $\mathcal{S}_1 \subseteq \mathcal{V}$. $\mathcal{G}$ is said to be strongly $r$-robust w.r.t. $\mathcal{S}_1$ if for any nonempty subset $\mathcal{S}_2 \subseteq \mathcal{V} \backslash \mathcal{S}_1$, $\mathcal{S}_2$ is r-reachable, where $r \in \mathbb{Z}_{>0}$.
\label{defstrongrrobust}
\end{Def}

\begin{figure*}[!t]
	\centering  
        \subfloat[]{
            \includegraphics[width=0.22\linewidth]{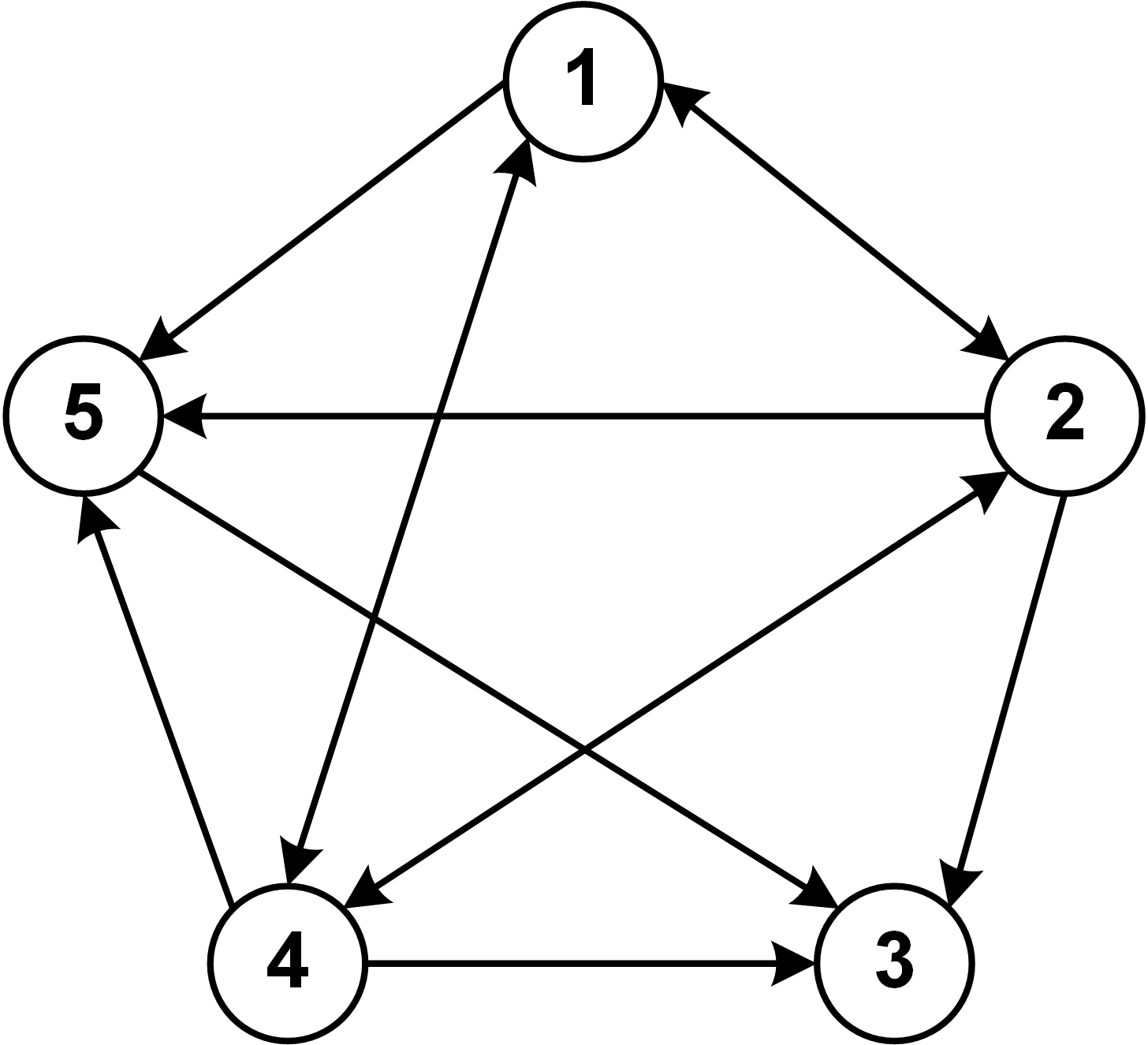}}
            \quad \quad
        \subfloat[]{
            \includegraphics[width=0.22\linewidth]{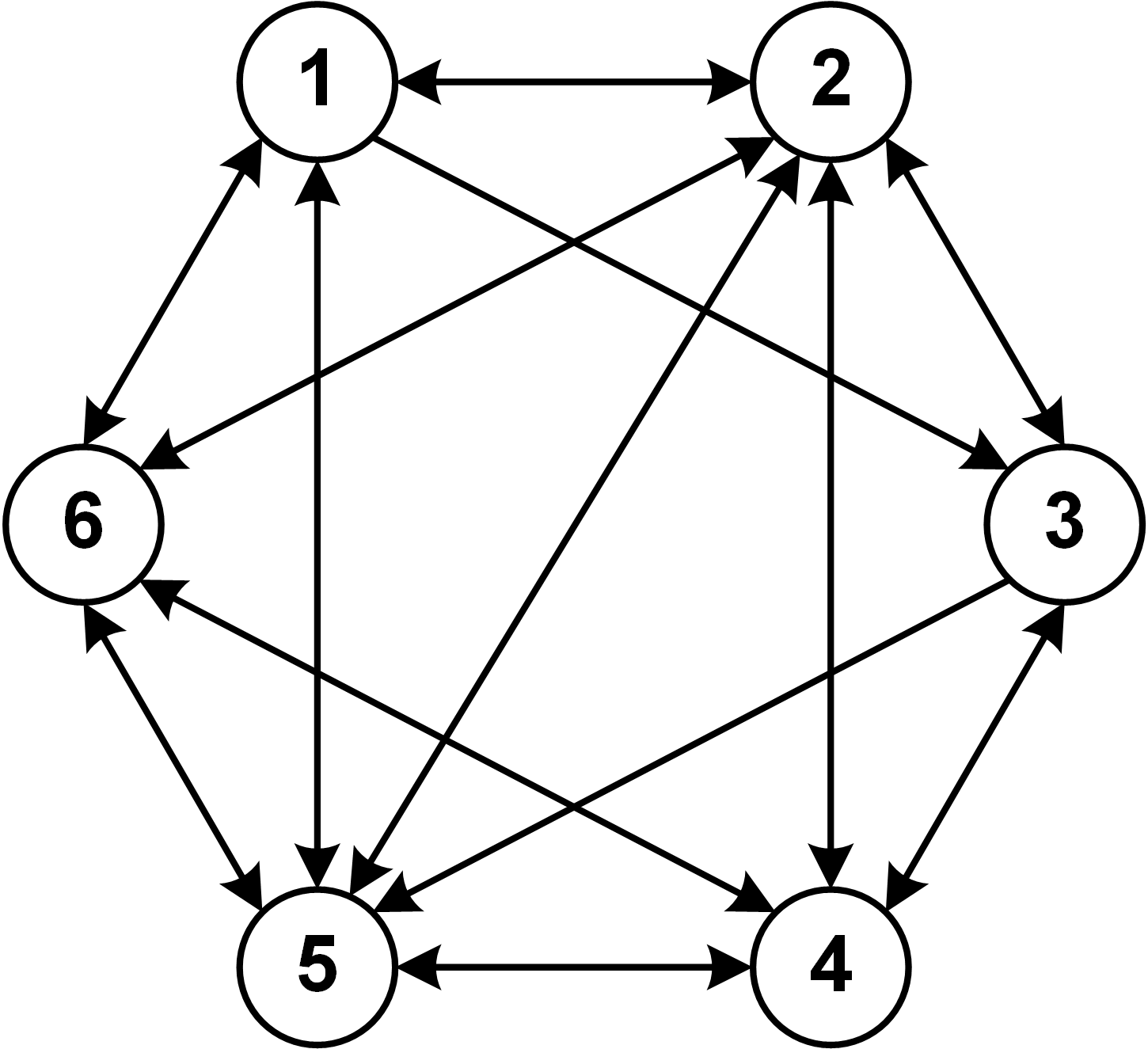}} 
            \quad \quad
        \subfloat[]{
            \includegraphics[width=0.22\linewidth]{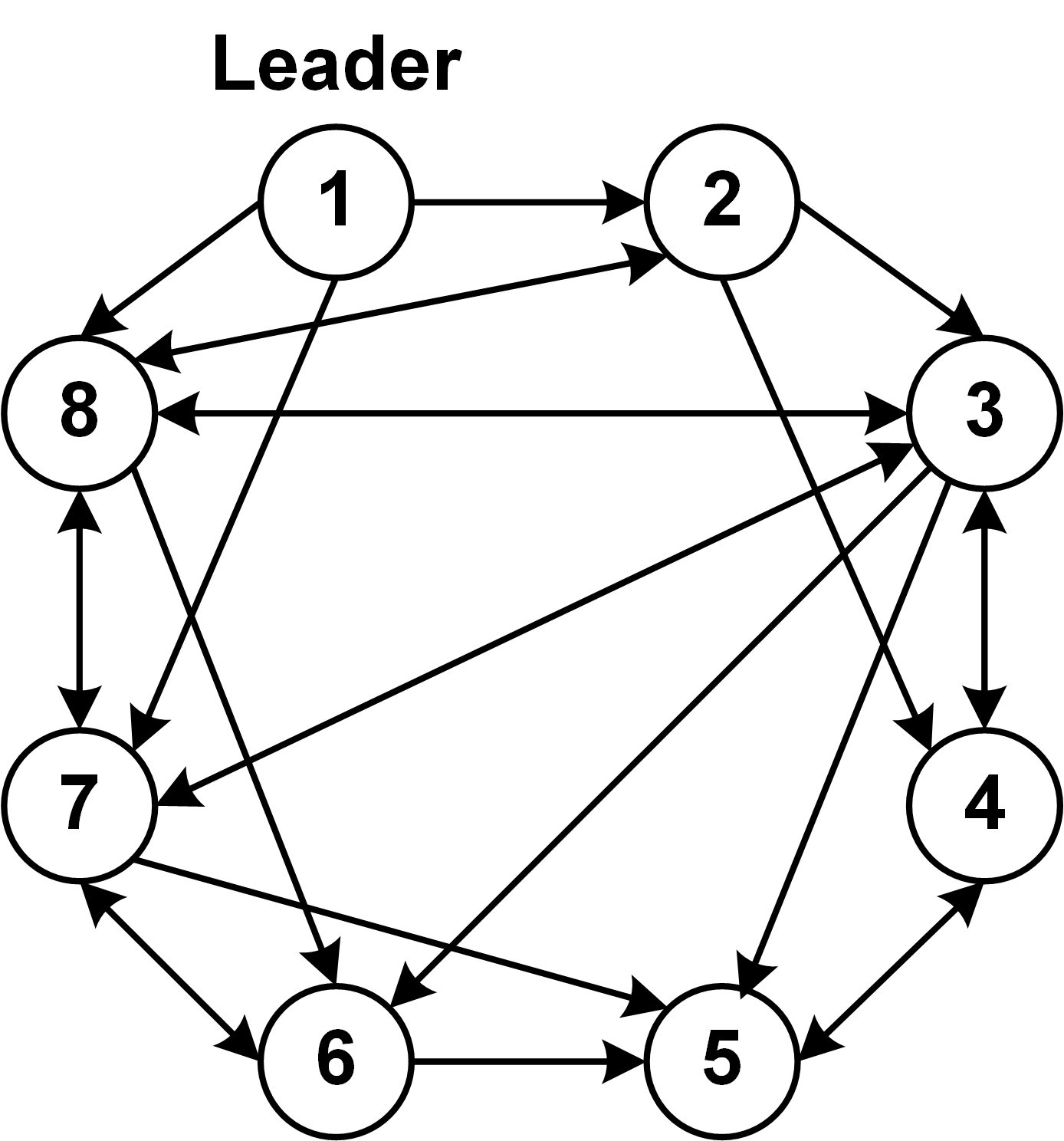}} 
	\caption{Illustrations of different robust network topologies: (a) (2,2)-robust graph, (b) 3-robust graph, and (c) strongly 3-robust graph.}
	\label{robustgraph}
\end{figure*}

Fig.~\ref{robustgraph} shows three different types of robust graphs, where agents are not filled with any color to distinguish their identities. This is because the graph robustness is merely associated with the communication links between agents. For $(r,s)$-robust and $r$-robust graphs, all agents are on equal footing and possess sufficient in-neighbors for state update, as illustrated in Figs.~\ref{robustgraph}(a) and \ref{robustgraph}(b). For a strongly $r$-robust graph, a certain number of agents are chosen as the leaders. Fig.~\ref{robustgraph}(c) provides one such example, where the digraph is strongly 3-robust w.r.t. $\mathcal{L}=\left\{ 1\right\}$.

Next, we will briefly discuss the relationship between these three graphs. For one thing, a $(r+s-1)$-robust graph is a sufficient condition to construct a $(r,s)$-robust graph. For another, consider an extreme case $\mathcal{S}_1= \varnothing$. In this case, the definition of strongly $r$-robust graph becomes the following: for any nonempty subset $\mathcal{S}_2 \in \mathcal{V}$, $\mathcal{S}_2$ is $r$-reachable, where $r \in \mathbb{Z}_{>0}$. By comparing it with Definition~\ref{defrrobust}, we can intuitively observe that a strongly $r$-robust graph is a sufficient condition for a $r$-robust graph in this particular case.


Note that the aforementioned concepts are established over time-invariant networks. In fact, numerous studies have extended the definitions of robust graphs from different perspectives (e.g., time-varying networks \cite{saldana2017resilient,usevitch2019resilient,wen2023joint}). The variants of these definitions will be examined later.

\section{Resilient Coordination Under Different Physical Structures}
\label{sec3}
This section will introduce several typical resilient strategies for addressing consensus problems in the presence of node injection attacks. These algorithms involve diverse system dynamics and structures. Specifically, we examine the following four cases: (i) single-integrator MAS \cite{dolev1982byzantine}, (ii) double-integrator MAS \cite{dibaji2015consensus,dibaji2017resilient}, (iii) general linear MAS \cite{bai2022resilient}, and (iv) leader-follower MAS \cite{usevitch2018resilient}.

\subsection{Single-Integrator MASs}
In \cite{dolev1982byzantine}, a purely local approach called \textit{approximate agreement} was proposed, where each normal agent neglects the possibly malicious information from its in-neighbors and updates its state based on safe information. In particular, they eliminate certain edges from in-neighbors with extremely large and small values at each iteration. It is a fault-tolerant algorithm since normal agents have no knowledge of the identities of malicious information. In cases of resilient consensus problems, this kind of strategy is widely acknowledged as the MSR algorithm \cite{kieckhafer1994reaching}. Subsequently, the seminal work \cite{leblanc2013resilient} developed a weighted MSR (W-MSR) algorithm to guarantee resilient consensus for single-integrator MASs. Consider a singe integrator MAS whose interactions are represented by the graph $\mathcal{G}=(\mathcal{V}, \mathcal{E})$. Each agent has a single-integrator structure given by $\dot{x}_i[t]=u_i[t], \ \ i \in \mathcal{V}$, where $x_i[t]$ is the state value and $u_i[t]$ is the control input.

In the absence of node injection attacks, the paper \cite{ren2005consensus} presented a discrete-time update scheme to make all normal agents in the MAS asymptotically converge to the consensus value, which is expressed as
\begin{equation*}
    x_i[t+1]=x_i[t]+\sum_{j \in \mathcal{V}_i^+[t]} a_{i j}[t]\left(x_j[t]-x_i[t]\right),
\end{equation*}
where the weights $a_{i j}[t]$ satisfy the following conditions:
\begin{enumerate}
    \item there exists a constant $\omega >0$ such that $a_{i j}[t] \geq \omega , \ \forall j \in \mathcal{V}_i^+[t], t \in \mathbb{Z}_{\geq 0}$;
    \item $a_{ij}[t]=0$ whenever $j \notin \mathcal{V}_i^+[t], t \in \mathbb{Z}_{\geq 0}$;
    \item $\sum_{j=1}^{n}a_{ij}[t]=1, \  \forall t \in \mathbb{Z}_{\geq 0}$.
\end{enumerate}

Under the $F$-total/local attack models, the work \cite{leblanc2013resilient} developed the W-MSR algorithm to defend against node injection attack and achieve resilient consensus in an asymptotic manner. The detailed procedure of the W-MSR algorithm is implemented as follows:

\begin{enumerate}
    \item \textit{(Collecting in-neighbors' information):} At each time step $t$, each normal agent $i \in \mathcal{N}$ broadcasts $x_i[t]$ to its out-neighbors, receives $x_j[t]$ from its in-neighbors $j \in \mathcal{V}_i^+[t]$, and sorts them in an ascending order.
    \item \textit{(Eliminating malicious states):} Compared with $x_i[t]$, agent~$i$ removes the $F$ smallest and largest values in the sorted list. If there are less than $F$ values strictly larger or smaller than $x_i[t]$, then all of the values that are strictly larger or smaller than $x_i[t]$ will be removed. The removal of these suspicious values is achieved through $a_{ij}[t]=0$. The state update for agent~$i$ will not utilize these removed data as they are considered malicious.
    \item \textit{(Updating local state):} Denote $\mathcal{R}_i^+[t]$ as the set of retained in-neighbors for agent~$i$. Then, the MAS adopts the following protocol for state update.
    \begin{equation}
    x_i[t+1]=x_i[t]+ \sum_{j \in \mathcal{R}_i^+[t]} a_{i j}[t]\left(x_j[t]-x_i[t]\right).
    \label{scheme2}
    \end{equation}
\end{enumerate}

It is worth noting that only the normal agents adhere to the W-MSR algorithm, whereas the state update of adversary agents may be manipulated by the attackers. To capture the property for $f$-fraction local models, the required parameter $F$ should be substituted by $F_i=\left\lfloor f\left|\mathcal{V}_i^+[t]\right| \right\rfloor$. In addition, the weight $a_{i j}[t]$ in (\ref{scheme2}) also satisfy the aforementioned conditions, but with $\mathcal{V}_i^+[t]$ substituted by $\mathcal{R}_i^+[t]$.


The main feature of the W-MSR algorithm lies in its lightweight. It does not require normal agents to know the overall network topology or the identities of non-neighbor agents. In the context of these advantages, a primary challenge for the W-MSR algorithm is to determine graph conditions that ensure normal agents to reach resilient consensus under different attack models. These conditions are closely associated with the parameter $F$, as we will see below.

The most fundamental investigation for resilient consensus is under the $F$-total malicious model, where the whole network topology is influenced by at most $F$ malicious agents. The following theorem provides the necessary and sufficient conditions for the normal agents to achieve resilient consensus under the $F$-total malicious models.
\begin{The}
\label{mainresultsingle}
   Consider a single-integrator MAS described by a time-invariant graph $\mathcal{G}=(\mathcal{V}, \mathcal{E})$. Suppose that the network satisfies the $F$-total malicious model and the normal agents execute the W-MSR algorithm for update. Then, resilient consensus is achieved if and only if the underlying network is $(F+1,F+1)$-robust. 
\end{The}

For the $F$-local malicious model, more malicious agents are allowed in the system since each normal agent may be influenced by at most $F$ malicious agents in its in-neighbor set. Consequently, the graph condition becomes more stringent compared to the $F$-total malicious model.


\begin{The}
\label{Theo22}
   Consider a single-integrator MAS described by a time-invariant graph $\mathcal{G}=(\mathcal{V}, \mathcal{E})$. Suppose that the network satisfies the $F$-local malicious model and the normal agents execute the W-MSR algorithm for state update.
   \begin{enumerate}
    \item A necessary condition for ensuring resilient consensus is that  $\mathcal{G}$ is $(F+1)$-robust.
    \item If the underlying network is $(2F+1)$-robust, then the MAS achieves resilient consensus.
\end{enumerate}
\end{The}

The third case involves the $f$-fraction local malicious model, where each agent may be influenced by at most a fraction $f$ of malicious agents in its in-neighbor set. By invoking the notion of $p$-fraction robust graph, the graph conditions for achieving resilient consensus under the $f$-fraction local malicious model are presented as follows:

\begin{The}
\label{Theo33}
   Consider a single-integrator MAS described by a time-invariant graph $\mathcal{G}=(\mathcal{V}, \mathcal{E})$. Suppose that the network satisfies the $f$-fraction local malicious model and the normal agents execute the W-MSR algorithm for update.
   \begin{enumerate}
    \item A necessary condition for achieving resilient consensus among normal agents is that the underlying network is $p^{\prime}$-robust, where $p^{\prime}>f$.
    \item If the underlying network is $p$-fraction robust, then the MAS achieves resilient consensus, where $p \in (2f,1]$.
\end{enumerate}
\end{The}

Note that the aforementioned results assume that the network is attacked by malicious agents. For the case of Byzantine attacks, the condition on network topologies are similar with those of malicious attacks under some attack models but slightly different under other models. Specifically, under the $F$-total Byzantine model, the graph condition for achieving resilient consensus is identical to Theorem~\ref{mainresultsingle}. Nevertheless, for the $F$-local Byzantine and $f$-fraction local Byzantine cases, the following corollary exhibits different graph conditions from Theorems~\ref{Theo22} and \ref{Theo33}.

\begin{Coro}
    Consider a single-integrator MAS described by $\mathcal{G}=(\mathcal{V}, \mathcal{E})$. Suppose that the normal agents execute the W-MSR algorithm for update. Under the $F$-local Byzantine model, resilient consensus is achieved if and only if $\mathcal{G}$ is $(F+1)$-robust. Under the $f$-fraction local Byzantine model, a necessary condition for achieving resilient consensus among normal agents is that the underlying network is $p^{\prime}$-fraction robust, where $p^{\prime} \geq f$, and resilient consensus is achieved if the underlying network is $p$-fraction robust, where $p>f$.
\end{Coro}

\subsection{Double-Integrator MASs}
In this subsection, we examine resilient consensus problems for double-integrator MASs \cite{dibaji2015consensus}. This kind of problem is motivated by autonomous mobile vehicles and robots \cite{zhang2009consentability,cao2010sampled}. Specifically, we consider agents possessing double-integrator dynamics and being subject to node injection attacks. In this context, we should focus on two key factors. Firstly, the adversary agents in the network may prevent the MAS from achieving consensus. Secondly, compared to the single-integrator case, system dynamics are more complicated, which requires consensus in both position and velocity.

Consider a network of agents with the double-integrator dynamics given by $\dot{x}_i(t)=v_i(t), \ \ \dot{v}_i(t)=u_i(t), \ \ i \in \mathcal{V}$, where $x_i(t)$ is the position\footnote{In order to distinguish different time domains, we use round brackets to represent the state in the continuous-time domain and square brackets to represent the state in the discrete-time domain.}, $v_i(t)$ is the velocity, and $u_i(t)$ is the control input.  Note that all these values are in the continuous-time domain.

Subsequently, the MAS is discretized by a sampling period $T$ and the update rule for the double-integrator MAS is presented as
\begin{equation}
\begin{aligned}
& x_i[k+1]=x_i[k]+T v_i[k]+\frac{T^2}{2} u_i[k], \\
& v_i[k+1]=v_i[k]+T u_i[k], i \in \mathcal{V},
\end{aligned}
\label{double-integrator dis}
\end{equation}
where $x_i[k] \in \mathbb{R}$, $v_i[k] \in \mathbb{R}$, and $u_i[k] \in \mathbb{R}$ denote the position, velocity and control input for agent $i$ at $t=kT$, respectively. 

The work \cite{dibaji2015consensus} aims to make all the normal agents in the MAS asymptotically reach consensus despite the influence of adversary agents. On this basis, the control input for the normal agents is designed considering agents' velocity and the relative error between the agents' position and the desired relative position. Therefore, the control protocol for normal agents is presented as
\begin{equation}
u_i[k]=-\alpha v_i[k]-\sum_{j \in \mathcal{R}_i^+[k]} a_{i j}[k]\left[\left(x_i[k]-\delta_i\right)-\left(x_j[k]-\delta_j\right)\right],
\label{control input double-integrator}
\end{equation}
where $\delta_i \in \mathbb{R}$ refers to the desired relative position for agent in a formation \cite{ren2011distributed}, and $\alpha$ is a positive scalar. A new state concerning the error from the desired position is further denoted as $\breve{x}_i[k]=x_i[k]-\delta_i$. In order to avoid the undesired oscillatory behavior \cite{qin2012sufficient}, the parameters $\alpha$ and sampling time $T$ should satisfy $1+\frac{T^2}{2} \leq \alpha T \leq 2-\frac{T^2}{2}$.

Instead of reaching resilient consensus defined in Definition~\ref{Problemerc}, the normal agents will endeavor to tackle the following double-integrator resilient consensus problem:
\begin{Def}[\textit{Double-integrator resilient consensus}]
\label{Prodi}
    Consider a graph $\mathcal{G}=(\mathcal{V}, \mathcal{E})$. Suppose that the network satisfies a certain threat model ($F$-total, $F$-local or $f$-fraction local). For any choice of initial position and velocity, determine graph conditions and design controllers such that the following conditions are satisfied.
\begin{itemize}
    \item Resilience condition: For each normal agent $i \in \mathcal{N}$ and for all $k \in \mathbb{Z}_{\geq 0}$, it holds $x_i[k] \in \mathcal{S}$, where
    \begin{equation*}
        \mathcal{S}= \left[\min _{i \in \mathcal{N}} \breve{x}_i(0)+\min _{i \in \mathcal{N}}\left\{0,\left(T-\frac{\alpha T^2}{2}\right) v_i(0)\right\}, \max _{i \in \mathcal{N}} \breve{x}_i(0)+\max _{i \in \mathcal{N}}\left\{0,\left(T-\frac{\alpha T^2}{2}\right) v_i(0)\right\} \right].
    \end{equation*}
    \item Consensus condition: For each normal agents $i \in \mathcal{N}$ and some constant $ b \in \mathcal{S}$, it holds $\lim_{k \rightarrow \infty}  \breve{x}_i[k] = b$ and $\lim_{k \rightarrow \infty} v_i[k]=0$.
\end{itemize}
\end{Def}

To solve the double-integrator resilient consensus problem, the work \cite{dibaji2015consensus} developed a \textit{Double-integrator Position-based Mean-Subsequence-Reduced} (DP-MSR) algorithm. The detailed procedures are implemented as follows.
\begin{enumerate}
    \item \textit{(Collecting in-neighbors' information):} At each time step $k$, each normal agent $i \in \mathcal{N}$ broadcasts $\breve{x}_i[k]$ to its out-neighbors, receives $\breve{x}_j[k]$ from its in-neighbors $j \in \mathcal{V}_i^+[k]$, and sorts $\breve{x}_j[k]-\breve{x}_i[k]$ in ascending order.
    \item \textit{(Eliminating malicious states):} Compared with zero, agent $i$ removes the $F$ smallest and largest values $\breve{x}_j[k]-\breve{x}_i[k]$ in the sorted list. If there are less than $F$ values strictly larger or smaller than zero, then all of the values that are strictly larger or smaller than zero will be removed. The removal of $\breve{x}_j[k]-\breve{x}_i[k]$ is achieved through $a_{ij}[k]=0$.
    \item \textit{(Updating local state):} Each normal agent applies the protocol (\ref{control input double-integrator}) for state update.
\end{enumerate}

With the DP-MSR algorithm, we are now ready to present the main results for double-integrator MASs.
\begin{The}
\label{mainresultdouble}
    Consider a double-integrator MAS. Suppose that the network satisfies the $F$-total malicious model and the normal agents execute the DP-MSR algorithm for update. Then, resilient consensus is achieved if and only if the underlying network is $(F+1,F+1)$-robust.
\end{The}

It was further revealed in \cite{dibaji2017resilient} that a $(2F+1)$-robust graph is a sufficient condition for the double-integrator MAS to achieve resilient consensus over networks with partial asynchrony and delay, when the system is subject to the $F$-local malicious attack. Note that only position values are processed trough the DP-MSR algorithm, whereas the velocities are directly utilized for state update. Consequently, the state that the double-integrator MAS eventually attains is called \textit{static consensus}, where the positions of all normal agents converge to a consensus value within the safety interval, and the velocities converge to zero, as shown in the consensus condition in Definition~\ref{Prodi}.

In addition, the paper \cite{bai2022resilient} designs a resilient consensus protocol
for a network of agents possessing general linear dynamics. The work \cite{oksuz2019resilient} studies the resilient consensus problem for nonlinear MASs, where a strictly increasing continuous nonlinear function is incorporated in the update rule to model the nonlinear dynamics. Both of them adopt the continuous-time MSR (CT-MSR) algorithm. This fact indicates that various system dynamics facilitate the development of MSR-type algorithms.

\subsection{Leader-Follower MASs}
\begin{figure*}[!t]
	\centering  
        \subfloat[Leaderless MAS.]{
            \includegraphics[height=0.275\textheight]{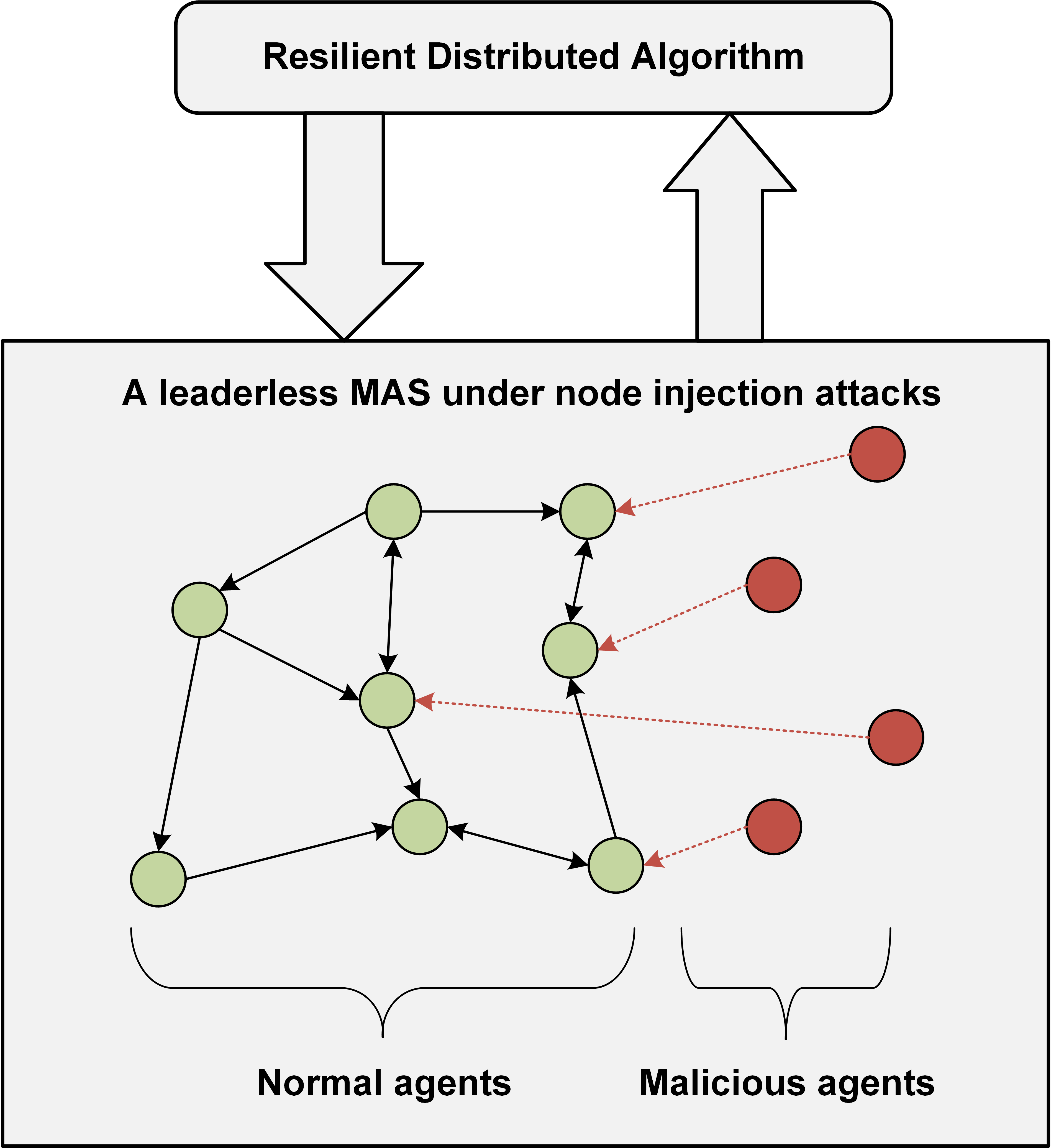}}
        \quad \quad
        \subfloat[Leader-follower MASs.]{
            \includegraphics[height=0.275\textheight]{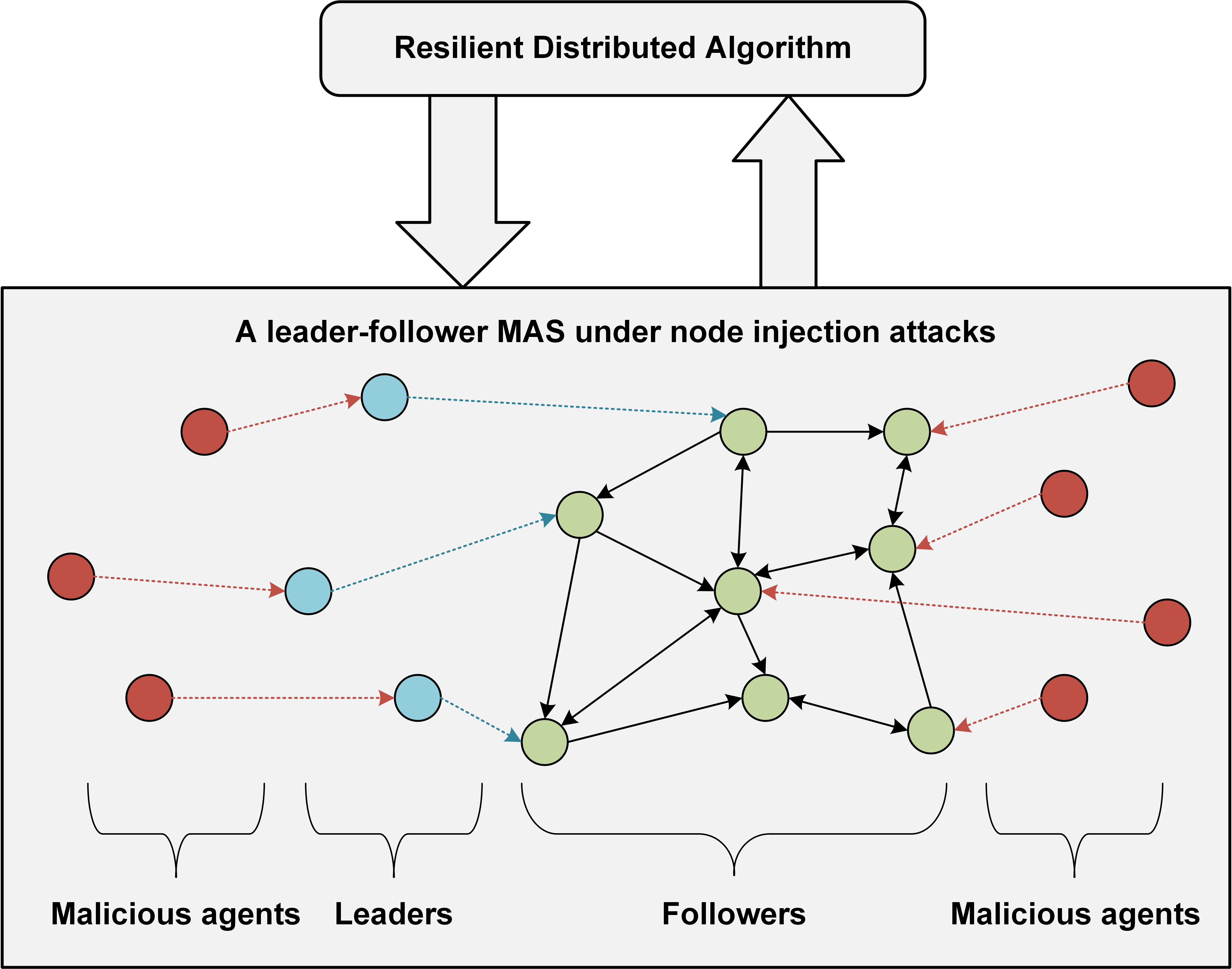}} 
	\caption{Scenarios of leaderless and leader-follower MASs under node injection attacks.}
	\label{attack1}
\end{figure*}

The aforementioned studies focus on resilient consensus problems under a leaderless structure. In a leader-follower MAS, the consensus problem under adversarial environments exhibits higher complexity since both the leaders and followers may be manipulated by attackers. In this case, stringent graph conditions are required, as illustrated in Fig.~\ref{attack1}. In this subsection, we examine the resilient consensus problem for leader-follower MASs with single-integrator dynamics and see how the W-MSR algorithm operates to defend against the node injection attack. Moreover, the notion of trusted node is introduced to relax the requirement for graph robustness.

Under a leader-follower structure, the node set $\mathcal{V}$ of the network $\mathcal{G}$ consists of $\mathcal{V} = \mathcal{L} \cup \mathcal{F}$, with $\mathcal{L}$ and $\mathcal{F}$ being the node sets of leaders and followers, respectively. By referring the concepts in node injection attacks, we further define $\mathcal{L}_n$ and $\mathcal{F}_n$ as the sets of normal leaders and normal followers, respectively. Each normal leader $l \in \mathcal{L}_n$ is able to send its state value to its out-neighbors at each time instant $t$. At the same time, each normal follower $i \in \mathcal{F}_n$ can receive state values from its in-neighbors and send its own state value to its out-neighbors at each time instant $t$.

In the context of leader-follower resilient consensus, the objective of normal leaders is to propagate a reference value (either static or dynamic). For normal followers, rather than solving the resilient consensus problem presented in Definition~\ref{Problemerc}, they will endeavor to tackle the following leader-follower resilient consensus problem:
\begin{Def}[\textit{Leader-follower resilient consensus}]
\label{Problemlfrs}
    Consider a leader-follower MAS described by a graph $\mathcal{G}=(\mathcal{V}, \mathcal{E})$. Suppose that the network satisfies a certain threat model ($F$-total, $F$-local or $f$-fraction local). For any choice of initial values, determine graph conditions and design controllers such that $ \lim _{t \rightarrow \infty} \left|x_i[t]-x_l[t] \right| = 0, \ \ \forall i \in \mathcal{F}_n,  \ \forall l \in \mathcal{L}_n$.
\end{Def}

Note that the leader-follower resilient consensus problem  merely poses a convergence requirement to the final state values of all normal followers, whereas the control inputs of leaders are designed individually. In fact, leader-follower MASs often exhibit deterministic behavior, meaning that the final consensus value is predictable.
The paper \cite{usevitch2018resilient} considers the static leader case. 
Benefited from the leader-follower structure, the most significant contribution in \cite{usevitch2018resilient} is that the final state values of normal followers converge to an arbitrary value, which is completely determined by the leaders. This property differs from the resilience condition presented in Definition~\ref{Problemerc} and renders resilient consensus more flexible and adaptable to various scenarios.

Based on Definition~\ref{defstrongrrobust}, the following theorem is presented to provide a sufficient condition on network topology for achieving leader-follower resilient consensus under the $F$-local malicious case.

\begin{The}
\label{mainresultnonlinear}
    Consider a leader-follower MAS described by $\mathcal{G}=(\mathcal{V}, \mathcal{E})$. Suppose that the network satisfies the $F$-local malicious model with $\left| \mathcal{L} \cap \mathcal{M} \right| \geq 0$ and the normal agents execute the W-MSR algorithm for update. Then, the leader-follower resilient consensus is achieved if the underlying network is strongly $(2F+1)$-robust w.r.t. $\mathcal{L}$.
\end{The}

\begin{table*}[!t]
\caption{Summary of Key References on Resilient Coordination Under Different Physical Structures.}
\label{Tab2}
\centering
\resizebox{0.95\linewidth}{!}{
\begin{tabular}{|c|c|c|c|c|c|c|}
\hline
References         & System Model                       & System structure            & Time Domain               & Resilient Algorithm    & Attack Model       & Graph Condition     \\ \hline
\multirow{3}{*}{\cite{leblanc2013resilient}} & \multirow{3}{*}{Single-integrator} & \multirow{3}{*}{Leaderless} & \multirow{3}{*}{Discrete} & \multirow{3}{*}{W-MSR} & $F$-total          & $(F+1,F+1)$-robust  \\ \cline{6-7} 
                   &                                    &                             &                           &                        & $F$-local          & $(2F+1)$-robust     \\ \cline{6-7} 
                   &                                    &                             &                           &                        & $f$-fraction local & $p$-fraction robust \\ \hline
\multirow{2}{*}{\cite{dibaji2017resilient2}} & \multirow{2}{*}{Double-integrator} & \multirow{2}{*}{Leaderless}      & \multirow{2}{*}{Discrete}   & \multirow{2}{*}{DP-MSR} & $F$-total      &$ (F+1,F+1)$-robust       \\ \cline{6-7} 
                         &                                    &                                  &                             &                         & $F$-local      & $(2F+1)$-robust          \\ \hline
\multirow{2}{*}{\cite{bai2022resilient}} & \multirow{2}{*}{General linear}    & \multirow{2}{*}{Leaderless}      & \multirow{2}{*}{Continuous} & \multirow{2}{*}{CT-MSR} & $F$-total      &$ (F+1,F+1)$-robust       \\ \cline{6-7} 
                         &                                    &                                  &                             &                         & $F$-local      & $(2F+1)$-robust          \\ \hline
\cite{oksuz2019resilient} &  Nonlinear & Leaderless & Continuous & CT-MSR & $F$-total & $(F+1)$-robust \\ \hline
\multirow{2}{*}{\cite{usevitch2018resilient}} & \multirow{2}{*}{Single-integrator} & \multirow{2}{*}{Leader-follower} & \multirow{2}{*}{Discrete}   & \multirow{2}{*}{W-MSR} & $F$-local      & Strongly$ (2F+1)$-robust \\ \cline{6-7} 
                         &                                    &                                  &                             &                         & $F$-local      & TLF robust with $F$            \\ \hline
\end{tabular}
}
\end{table*}

An alternative method to achieve the leader-follower resilient consensus is to incorporate trusted nodes, which are defined as below. 
\begin{Def}[\cite{abbas2014resilient}]
An agent is said to be trusted if it follows the prescribed rule and cannot be compromised by malicious attacks. We denote the set of trusted agents as $\mathcal{T} \subseteq \mathcal{N}$.
\end{Def}

Trusted nodes are assumed sufficiently secure such that they cannot be compromised by malicious attacks. The work \cite{usevitch2018resilient} incorporated them into the leader-follower resilient consensus problem that defines a novel robust graph as follows.
\begin{Def}[\textit{trusted leader-follower robust graph}]
    Consider a time-invariant graph $\mathcal{G}=(\mathcal{V}, \mathcal{E})$ and a nonempty subset $\mathcal{S}_1 \subseteq \mathcal{V}$. $\mathcal{G}$ is said to be trusted leader-follower (TLF) robust with $F$ if for any nonempty subset $\mathcal{S}_2 \subseteq \mathcal{V} \backslash \mathcal{S}_1$, at least one of the following condition is satisfied, where $F \in \mathbb{Z}_{\geq 0}$.
    \begin{itemize}
        \item There exists $i \in \mathcal{S}_2$ with at least $F+1$ in-neighbors from $\mathcal{S}_1$, i.e., $\left| \mathcal{V}_i^+ \cap \mathcal{S}_1 \right| \geq F+1$;
        \item $\mathcal{S}_2$ is $(2F+1)$-reachable.
    \end{itemize}
\end{Def}

Leveraging this definition, the authors in \cite{usevitch2018resilient} further derived the following result.
\begin{The}
\label{mainresultleaderfollowertrust}
    Consider a leader-follower MAS described by $\mathcal{G}$. Suppose that the network satisfies the $F$-local malicious model with $\left| \mathcal{L} \cap \mathcal{M} \right| = 0$ and the normal agents execute the W-MSR algorithm for update. Then, leader-follower resilient consensus is achieved if the underlying network is strongly TLF robust with parameter $F$.
\end{The}
\begin{Rem}
    It is worth mentioning that for a graph to be strongly $(2F +1)$-robust w.r.t. $\mathcal{L}$, it should hold that $\left| \mathcal{L} \right| \geq 2F + 1$. After the trusted nodes are incorporated into the network, this lower bound on the number of leaders is relaxed to some extent.  More specifically, under a TLF robust graph, the minimum number of leaders required reduces from $2F+1$ to $F+1$. In addition, the graph condition for achieving resilient consensus under leader-follower structures is more stringent compared to leaderless resilient consensus, which also indicates the necessity to reduce graph requirements. 
\end{Rem}

\begin{Rem}
    The choice between leader-follower and leaderless consensus depends on the specific requirements and constraints of the distributed system. Some systems may benefit from the advantages of leader-follower consensus, while others may prefer the fault-tolerance and decentralized nature of leaderless consensus.
\end{Rem}

TABLE~\ref{Tab2} provides a summary of the key references reviewed in this section. Note that ``Graph Condition'' refers to a sufficient condition for the MAS to achieve resilient consensus.

\section{Resilient Coordination over Time-Varying Network Topologies}
\label{sec4}
In addition to examining the resilient coordination problems with various system structures and dynamics, effect of time-varying networks \cite{sun2009consensus} is contemplated as a significant part to handle the situations of constrained cyber capacity and physical hindrances during the information interaction and transmission. As discussed in the previous subsection, resilient consensus is assured if the network satisfies specific robust requirements. However, the multi-agent network may not be robust at each time step and its communication link may be activated selectively in some cases. Under these constraints and assumptions, how to design control strategies and determine graph conditions such that the MAS achieves resilient consensus becomes a challenging problem.

Compared to time-invariant networks, time-varying networks are more common which relax the requirement that the digraph should satisfy certain graph conditions at each time step. Currently, there are several studies that have investigated resilient consensus problems in time-varying networks \cite{leblanc2013resilient,saldana2017resilient,huang2019resilient,usevitch2019resilient,wen2023joint}. In the presence of $F$-local malicious model, the seminal work \cite{leblanc2013resilient} presents the following corollary, which lays a foundation for analyzing resilient consensus problems in time-varying network topologies.
\begin{Coro}
   Consider a single-integrator MAS described by a time-varying graph $\mathcal{G}[t]=(\mathcal{V}, \mathcal{E}[t])$. Suppose that the network satisfies the $F$-local malicious model and the normal agents execute the W-MSR algorithm for update. Denote $\{t_k\}$ as the set of time instants when $\mathcal{G}[t]$ is $(2F +1)$-robust. Then, resilient consensus is achieved if $| \{t_k\} |=\infty$ and $|t_{k+1}-t_k|\leq c, \ \forall k \in \mathbb{Z}_{\geq 0}$, where $c \in \mathbb{Z}_{>0}$.
   \label{Coro time-varying}
\end{Coro}

\subsection{$(T,r)$-Robust Graph}
Motivated by Corollary~\ref{Coro time-varying}, the work \cite{saldana2017resilient} developed a sliding window W-MSR (SW-MSR) algorithm to guarantee resilience for networks of agents in time-varying graphs. Particularly, the authors in \cite{saldana2017resilient} presented a time-varying version of $r$-robust graphs by introducing the time window $T$.

\begin{Def}[\textit{$(T,r)$-robust graph}]
    Consider a time-varying graph $\mathcal{G}[t]=(\mathcal{V}, \mathcal{E}[t])$. $\mathcal{G}[t]$ is said to be $(T,r)$-robust if $\mathcal{G}^T[t]$ satisfies the conditions of an $r$-robust graph $\forall t \geq T$, where $T,r,t \in \mathbb{Z}_{>0}$ and $\mathcal{G}^T[t]=\cup_{\tau=0}^T \mathcal{G}[t-\tau]$.
    \label{deftrrobust}
\end{Def}

Leveraging this property, agents in the MAS are not required to establish robust graphs at each time step. Instead, they form such topologies jointly over finite intervals of time window $T$. To this end, the W-MSR algorithm is also extended to a time-varying version, called the \textit{Sliding Weighted Mean-Subsequence-Reduced} (SW-MSR) algorithm. The main steps are presented as follows.
\begin{enumerate}
    \item \textit{(Collecting in-neighbors' information):} Consider a window with duration $T$ steps. At each time step $t$, each normal agent $i \in \mathcal{N}$ receives $\{x_j[t-\tau_{ij}]| j \in \mathcal{V}_i^{T}[t]\}$ from its in-neighbors $j \in \mathcal{V}_i^T[t]$ and sorts them in ascending order, where $\mathcal{V}_i^{T}[t]=\cup_{\tau=t-T}^{t}\mathcal{V}_i^+[\tau]$, $\tau_{i j}[t]=\max \left(\left\{\tau \in[0, T] \mid j \in \mathcal{V}_i^+[t-\tau]\right\}\right), \forall j \in \mathcal{V}_i^T[t]$.
    \item \textit{(Eliminating malicious states):} Compared to $x_i[t]$, agent $i$ removes the $F$ smallest and largest values in the sorted list. If there are less than $F$ values strictly larger or smaller than $x_i[t]$, then all of the values that are strictly larger or smaller than $x_i[t]$ will be removed. The removal of values is achieved through $a_{ij}[t]=0$. The state update for agent $i$ will not utilize these removed data as they are considered suspicious.
    \item \textit{(Updating local state):} Denote $\mathcal{R}_i^+[t]$ as the set of retained in-neighbor values within the time window $T$ for agent $i$. Then, the MAS adopts the following protocol for state update.
    \begin{equation}
    x_i[t+1]=x_i[t]+ \sum_{j \in \mathcal{R}_i^+[t]} a_{i j}[t]\left(x_j[t-\tau_{ij}[t]]-x_i[t]\right).
    \label{shceme timevarying}
    \end{equation}
\end{enumerate}

If $T=0$, the SW-MSR algorithm degenerates into the W-MSR algorithm. Intuitively, the SW-MSR algorithm can be considered an extension of the W-MSR algorithm to graphs with time-varying edge sets. In comparison to the W-MSR algorithm, the most significant modification of the SW-MSR algorithm is that it incorporates all received values within the time window $T$. These values are originated from the temporal information remained by each agent during the reception of messages. Subsequently, the received values go through the screening process and the retained ones will be utilized for state update. Although the introduction of the time window $T$ may raise higher demands on the information storage capacity for agents, it relaxes the requirement that the network should satisfy certain graph conditions at each time step. Overall, this modification is more advantageous that helps the MAS achieve resilient consensus over time-varying networks.

In the following part, we provide a sufficient condition for the normal agents to achieve resilient consensus over time-varying networks, when the system is subject to $F$-local malicious model.

\begin{The}
   Consider a leaderless MAS described by a time-varying graph $\mathcal{G}[t]=(\mathcal{V}, \mathcal{E}[t])$. Suppose that the network satisfies the $F$-local malicious model and the normal agents execute the W-MSR algorithm for update. Then, resilient consensus is achieved if the underlying network is $(T,2F+1)$-robust. 
\end{The}

\subsection{Strongly $(T,t_0,r)$-Robust Graph}

Under a leader-follower structure, the paper \cite{usevitch2019resilient} investigated the resilient consensus problem in time-varying networks, where the normal leaders aim to propagate a reference value, and the normal followers seek to converge to this value. $(T,r,t_0)$-robust graph is defined as follows \cite{usevitch2019resilient}.
\begin{Def}[\textit{Strongly $(T,t_0,r)$-robust graph}]
Consider a time-varying graph $\mathcal{G}[t]=(\mathcal{V}, \mathcal{E}[t])$. $\mathcal{G}[t]$ is said to be strongly $(T,t_0,r)$-robust w.r.t. a nonempty set $\mathcal{S}_1 \subseteq \mathcal{V}$ if $\mathcal{G}^T[t]$ is strongly $r$-robust w.r.t. $\mathcal{S}_1$, where $\mathcal{G}^T[t]=\cup_{\tau=0}^T \mathcal{G}[t-\tau]$, $T,r,t \in \mathbb{Z}_{>0}$, $t_0 \in \mathbb{Z}$ and $t \geq t_0+T $.
\label{defstronglyttr}
\end{Def}

We should point out that though the definition of strongly $(T,t_0,r)$-robust graph is quite similar to Definition~\ref{deftrrobust}, the former is built on a leader-follower structure and involves an initial time $t_0$, so that the graph robustness does not need to be satisfied from time step zero. In addition, the introduction of the time window $T$ relaxes the requirement that the network should satisfy certain graph robustness conditions at each time step. Instead, only the union of $\mathcal{G}[t]$ over the last $T$ time steps are required to be strongly robust.

As stated earlier, the objective of the leaders is to propagate a reference value, thus the leaders follow the same static input as $x_l[t]= C, \ \forall t \geq t_0, \ \ \forall l \in \mathcal{L}_n$, where $C$ is a constant and $t_0$ is the initial time instant.

For the normal followers, they adhere to the SW-MSR algorithm and update their state values according to (\ref{shceme timevarying}). Nevertheless, there may exist some adversary agents in the network that deviate from the prescribed protocols. Instead, they apply some other rules for state update and disseminate malicious information to all of their out-neighbors. Such misbehavior may diminish the effectiveness of control strategies and disrupt system security. The following theorem provides a sufficient condition for the MAS to handle the leader-follower consensus problem under adversarial environments.

\begin{The}
   Consider a leader-follower MAS described by a time-varying graph $\mathcal{G}[t]=(\mathcal{V}, \mathcal{E}[t])$. Suppose that the network satisfies the $F$-local malicious model and the normal agents execute the SW-MSR algorithm for state update. Then, resilient consensus is achieved among the normal followers if $\mathcal{G}[t]$ is strongly $(T,t_0,2F+1)$-robust w.r.t. $\mathcal{L}$.
\end{The}

\subsection{Jointly $r$-Robust Graph}

Another way to extend the network topology to the time-varying version is through the notion of joint connectivity \cite{shi2009global,zhang2010general}. In \cite{wen2023joint}, the authors combined the graph robustness of the time-invariant network topology \cite{leblanc2013resilient} with the time-varying network jointly containing a directed spanning tree \cite{ren2005consensus}, thereby developing a novel notion of \textit{joint robustness} to characterize the graph robustness property for reaching resilient consensus of MASs over time-varying network topologies. The proposed notion allows for a significant reduction in communication overheads, hence creating potential chances for the practical implementation of distributed resilient consensus. The notions of \textit{joint reachability} and \textit{joint robustness} are defined as follows.

\begin{Def}[\textit{Jointly $r$-reachable set}]
    Consider a time-varying digraph $\mathcal{G}[t]=(\mathcal{V}, \mathcal{E}[t])$ and a nonempty subset $\mathcal{S} \subseteq \mathcal{V}$. $\mathcal{S}$ is said to be jointly $r$-reachable if there exists an infinite sequence of uniformly bounded and non-overlapping time intervals $[t_h,t_{h+1})$ such that in each time interval $[t_h,t_{h+1})$, there exist a time step $t_j \in [t_h,t_{h+1})$ and an agent $i_j \in \mathcal{S}$ such that $|\mathcal{V}_{i_j}^{+}[t_j]\backslash \mathcal{S} | \geq r$, where $r \in \mathbb{Z}_{>0}$, $t_h=k_h T, \ h=1,2,\cdots$, and $k_1=0$.
\end{Def}

\begin{Def}[\textit{Jointly $r$-robust graph}]
    Consider a time-varying graph $\mathcal{G}[t]=(\mathcal{V}, \mathcal{E}[t])$. $\mathcal{G}[t]$ is said to be jointly $r$-robust for each pair of nonempty, disjoint subsets $\mathcal{S}_1,\mathcal{S}_2 \subseteq \mathcal{V}$, at least one of them is jointly $r$-reachable, where $r \in \mathbb{Z}_{>0}$.
    \label{defjointlyrrobust}
\end{Def}

\begin{figure*}[!t]
    \centering
    \includegraphics[width=0.75\linewidth]{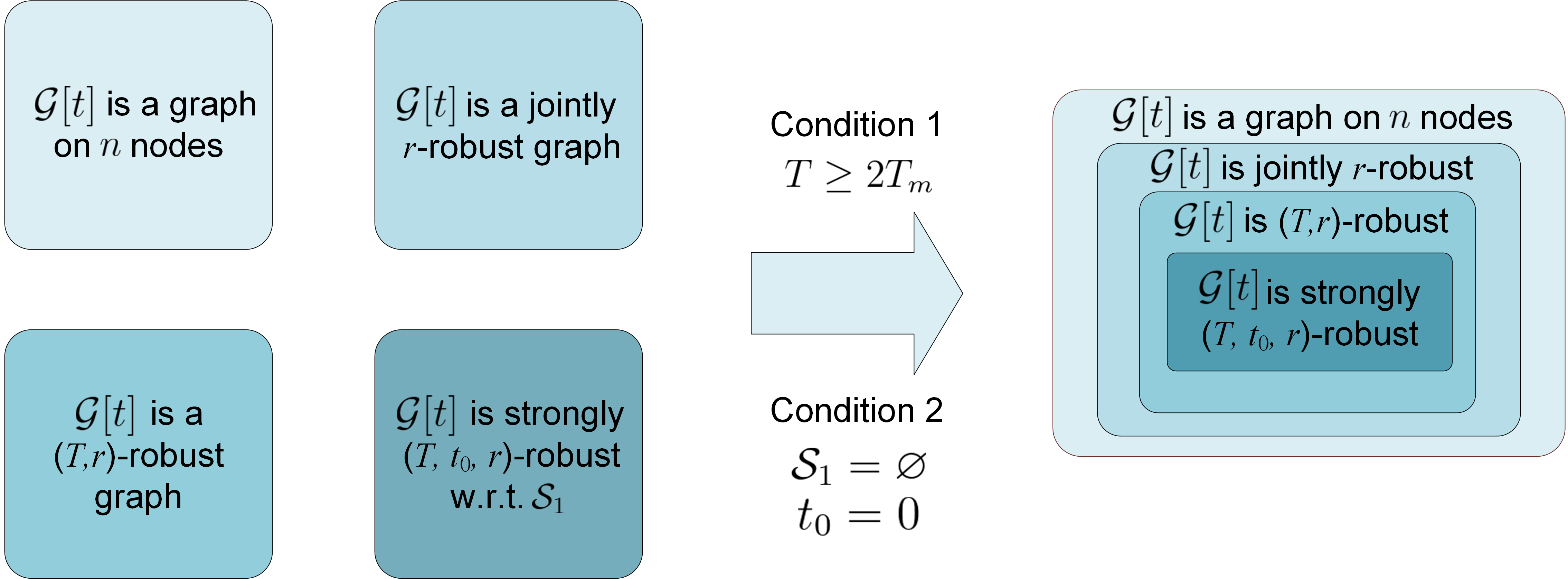}
    \caption{The relationship between three time-varying robust graphs.}
    \label{rela}
\end{figure*}

Before stating the graph condition for achieving time-varying resilient consensus utilizing the proposed notion of graph robustness, we explore the connection and difference between three robustness concepts in Definition~\ref{deftrrobust}, Definition~\ref{defstronglyttr}, and Definition~\ref{defjointlyrrobust}. The relationship between these three time-varying robust graphs is illustrated in Fig.~\ref{rela}. If the time-varying network $\mathcal{G}[t]$ is jointly $r$-robust, then there exists a $T$ such that $\mathcal{G}[t]$ is $(T,r)$-robust. This fact implies that the joint $r$-robust graph is a sufficient condition for the $(T,r)$-robust graph by choosing a sufficiently large $T$. Particularly, we can choose $T \geq 2T_m$, where $T_m$ represents the maximum duration of time periods $[t_h,t_{h+1})$ in Definition~\ref{defjointlyrrobust}. Furthermore, if we assume $\mathcal{S}_1 = \varnothing, t_0=0$, we will find that the $(T,t_0,r)$-robust graph is a special case of the $(T,r)$-robust graph. To better describe the relationship between three time-varying robust graphs when certain conditions are satisfied, a hierarchical containment diagram is depicted in Fig.~\ref{rela}.

Now, we are ready to give the necessary and sufficient conditions to achieve time-varying resilient consensus under the $F$-local malicious model.
\begin{The}
   Consider a leaderless MAS described by a time-varying graph $\mathcal{G}[t]=(\mathcal{V}, \mathcal{E}[t])$. Suppose that the network satisfies the $F$-local malicious model and the normal agents execute the W-MSR algorithm for state update.
   \begin{enumerate}
    \item A necessary condition for achieving resilient consensus among the normal agents is that $\mathcal{G}[t]$ is jointly $(F+1)$-robust.
    \item If $\mathcal{G}[t]$ is jointly $(2F+1)$-robust, then the MAS achieves resilient consensus.
\end{enumerate}
\end{The}

It was further revealed in \cite{wen2023joint} that joint $(2F+1)$-robustness is the necessary and sufficient condition for the MAS to achieve resilient consensus over time-varying networks, when the system is subject to the $F$-total malicious attack.

\begin{Rem}
    Among these studies on time-varying networks, one of their common characteristics is that they extend the notion of $r$-robust graph to different time-varying versions. Nevertheless, these works serve for different system structures, where agents need to execute different resilient algorithms, and the conditions on network topology are also different. Furthermore, as shown in Fig.~\ref{rela}, the relationship between the three robust graphs is generally independent. In some special cases, however, they exhibit an inclusion relationship. Therefore, how to find a uniform notion to generalize the resilient coordination in time-varying graphs still remains an open question. Alternatively, changing the weights $a_{ij}[t]$ periodically is another way to investigate resilient coordination problems over time-varying network topologies.
\end{Rem}

TABLE~\ref{Tab3} provides a summary of the key references on resilient coordination over different time-varying network topologies.

\begin{table*}[!t]
\caption{Summary of Key References on Resilient Coordination over Different Time-Varying Graphs.}
\label{Tab3}
\centering
\resizebox{1\linewidth}{!}{
\begin{tabular}{|c|c|c|c|c|c|c|}
\hline
References         & Network Topology                                                                    & System Model                                                                                       & System Structure            & Resilient Algorithm                                                         & Attack Model & Graph Condition            \\ \hline
\cite{saldana2017resilient}                  & \begin{tabular}[c]{@{}c@{}}$(T,r)$-robust\\ graph\end{tabular}                      & Single-integrator                                                                                  & Leaderless                  & SW-MSR                                                                      & $F$-local    & $(T,2F+1)$-robust          \\ \hline
\cite{usevitch2019resilient}                  & \begin{tabular}[c]{@{}c@{}}Strongly $(T,t_0,r)$-robust \\ graph\end{tabular}        & Single-integrator                                                                                  & Leader-follower             & SW-MSR                                                                      & $F$-local    & $(T,t_0,2F+1)$-robust      \\ \hline
\multirow{2}{*}{ \cite{wen2023joint} } & \multirow{2}{*}{\begin{tabular}[c]{@{}c@{}}Jointly $r$-robust\\ graph\end{tabular}} & \multirow{2}{*}{\begin{tabular}[c]{@{}c@{}}Single-integrator and\\ double-integrator\end{tabular}} & \multirow{2}{*}{Leaderless} & \multirow{2}{*}{\begin{tabular}[c]{@{}c@{}}W-MSR and\\ DP-MSR\end{tabular}} & $F$-total    & Jointly $(F+1,F+1)$-robust \\ \cline{6-7} 
                   &                                                                                     &                                                                                                    &                             &                                                                             & $F$-local    & Jointly $(2F+1)$-robust    \\ \hline
\end{tabular}
}
\end{table*}

\section{Resilient Coordination with Different Communication Mechanisms}
\label{sec5}
Currently, numerous resilient consensus algorithms are executed in a continuous-time or time-triggered communication mode. It means that frequent information interaction and transmission between agents are required to achieve consensus for the MAS. In hostile environments, the communication burden will become heavier since additional resilient algorithms should be considered to overcome the influence of malicious attacks, and more communication resources will be consumed at each time step. To mitigate the heavy communication burden, we introduce three other communication mechanisms besides the time-triggered pattern, namely the event-triggered, self-triggered, and quantized mechanisms. All of them are committed to reducing the communication frequency between agents and consuming less communication resources, but they show differences in the communication principle and controller design.

\subsection{Event-Triggered Communication Mechanism }

The event-triggered communication is a widely-used mechanism in distributed systems \cite{dimarogonas2011distributed,hu2015consensus,sun2018event,yi2018dynamic,li2020adaptive} to facilitate the exchange of information or messages between different components or processes based on specific events or conditions. In this communication model, messages are sent or received when predefined events occur, rather than on a fixed schedule or at regular intervals. This approach is particularly useful for optimizing system resources and reducing unnecessary communication overhead. Thus, extensive research has been conducted on designing event-based resilient coordination protocols \cite{wang2019resilient2,zegers2019event,wang2019resilient,wu2021resilient}. In \cite{wang2019resilient}, the authors developed an \textit{Event-based Mean-Subsequence-Reduced} (E-MSR) algorithm, which can be regarded as an asynchronous version of the W-MSR algorithm.

In order to implement the E-MSR algorithm, an auxiliary variable $\hat{x}_i[t] \in \mathbb{R}$ is introduced and defined as
\begin{equation*}
    \hat{x}_i[t]=x_i[t_l^i], \quad k \in [t_l^i,t_{l+1}^i),
\end{equation*}
where $\hat{x}_i[t]$ denotes the state value sent by agent $i$ at the last communication instant and $t_0^i,t_1^i,\ldots$ are the communication instants of agent $i$. These communication instants are determined by an event-triggered function (ETF), which is designed as 
\begin{equation}
f_i[t]=\left| e_i[t] \right|-\left(c_0+c_1 \mathrm{e}^{-\alpha t}\right),
\label{ETF}
\end{equation}
where $e_i[t]=x_i[t+1]-\hat{x}_i[t]$ is the relative error between the updated state $x_i[t+1]$ and the auxiliary variable $\hat{x}_i[t]$, $c_0$, $c_1$, and $\alpha$ are positive scalars, and the function $c_0+c_1 \mathrm{e}^{-\alpha t}$ denotes a threshold. If $f_i[t] > 0$, agent $i$ will substitute its auxiliary variable with its current state and transmit its current state to its in-neighbors.

Compared to the W-MSR algorithm, the main steps of the E-MSR algorithm are slightly modified and an additional step is involved to update the auxiliary variable according to the ETF. The detailed procedure of the E-MSR algorithm is presented as follows.
\begin{enumerate}
    \item \textit{(Collecting in-neighbors' information):} At each time step $k$, each normal agent $i \in \mathcal{N}$ broadcasts $\hat{x}_i[t]$ to its out-neighbors, receives $\hat{x}_j[t]$ from its in-neighbors $j \in \mathcal{V}_i^+[t]$, and sorts them in ascending order.
    \item \textit{(Eliminating malicious states):} Compared to $x_i[t]$, agent $i$ removes the $F$ smallest and largest values in the sorted list. If there are less than $F$ values strictly larger or smaller than $x_i[t]$, then all of the values that are strictly larger or smaller than $x_i[t]$ will be removed. The removal of values is achieved through $a_{ij}[t]=0$.
    \item \textit{(Updating local states):} Denote $\mathcal{R}_i^+[t]$ as the set of retained auxiliary variables for agent $i$. Then, the MAS adopts the following protocol for state update.
    \begin{equation}
    x_i[t+1]=x_i[t]+ \sum_{j \in \mathcal{R}_i^+[t]} a_{i j}[t]\left(\hat{x}_j[t]-x_i[t]\right).
    \label{schemeevent}
    \end{equation}
    \item \textit{(Updating auxiliary variables):}  Agent $i$ checks the positivity of the ETF (\ref{ETF}), then updates its auxiliary variable according to
       \begin{equation*}
           \hat{x}_i[t+1]=
           \begin{cases}
           x_i[t+1],&{\text{if}}\ \ f_i[t] > 0,\\
           {\hat{x}_i[t],}&{\text{otherwise.}}
            \end{cases}
       \end{equation*}
\end{enumerate}

Note that the threshold in (\ref{ETF}) consists of a positive constant term $c_0$ and an exponential term $c_1 \mathrm{e}^{-\alpha t}$. The relative error $e_i[t]$ will not converge to zero due to the influence of $c_0$. Consequently, it is difficult for the MAS to achieve (exact) resilient consensus presented in Definition~\ref{Problemerc}. Instead, the normal agents will endeavor to tackle the approximate resilient consensus problem. The following theorem not only provides the graph condition for the MAS to achieve approximate resilient consensus under the $F$-total malicious model, but also gives the error level to be accomplished by the normal agents despite the influence of malicious agents. 
\begin{The}
\label{resultevent}
    Consider a time-invariant graph $\mathcal{G}=(\mathcal{V}, \mathcal{E})$. Suppose that the network satisfies the $F$-total malicious model and the normal agents execute the E-MSR algorithm for update. Then, resilient consensus at an error level $c$ is achieved if and only if the underlying network is $(F +1, F +1)$-robust. Furthermore, the error level $c$ is accomplished if the parameter $c_0$ in the ETF (\ref{ETF}) satisfies $c_0 \leq \frac{\omega^{n} c}{4 n}$.
\end{The}
\begin{Rem}
    Although the event-triggered mechanism is able to reduce the communication burden to some extent, the threshold associated with the state error becomes smaller as the MAS approaches consensus, potentially resulting in more frequent triggers. Thus, it is desirable to adjust the threshold in the ETC dynamically and develop dynamic triggering schemes \cite{girard2014dynamic,yi2018dynamic,mishra2023event}. In addition, one disadvantage of the event-triggered mechanism is that it requires each normal agent to frequently monitor its own state and listen to its in-neighbors. This requirement can be removed if each normal agent determines its next triggering time at the current triggering time, which is also known as the self-triggered mechanism. This communication mechanism will be introduced in detail in the next subsection.
\end{Rem}

\subsection{Self-Triggered Communication Mechanism}

Self-triggered communication is a mechanism in which the MAS decides when and how to send or receive messages based on its internal state or predefined rules, without relying on external events or triggers \cite{dimarogonas2010distributed,fan2015self,hu2016output}. Unlike the event-triggered communication which responds to specific events or conditions, self-triggered communication is initiated by the system itself, often according to a set schedule or criteria. In adversarial environments, this characteristic is advantageous since adversary agents may make unnecessary communications and broadcast malicious information to other normal agents. In this subsection, we discuss a self-triggered control strategy and present how it is accomplished based on the ternary control \cite{matsume2021resilient}.

Consider a single-integrator MAS with ternary variables. In addition to state $x_i(t)$ and input $u_i(t)$, each agent $i \in \mathcal{V}$ possesses a local clock $\theta_i(t)$, which determines the communication and update events. The update of these three variables satisfy $\dot{x}_i(t)=u_i(t), \dot{u}_i(t)=0, \dot{\theta}_i(t)=-1$. However, this update is not applicable when $\theta_i(t)=0$. At these self-triggered times $\{ t_m^i\}_{m \in \mathbb{Z}_{>0}}$, the state update for agent $i$ adheres to the following ternary protocol:
\begin{equation}
u_i\left( t_m^i \right)= \begin{cases}\operatorname{sign}\left(f_{\varepsilon}\left(\operatorname{ave}_i(t)\right)\right) & \text { if } i \in \mathcal{U}(t) \\
u_i(t) & \text { otherwise }\end{cases} \\, \quad \theta_i\left( t_m^i \right)= \begin{cases}\max \left\{\left|\operatorname{ave}_i(t)\right|, \varepsilon\right\} & \text { if } i \in \mathcal{U}(t) \\
\theta_i(t) & \text { otherwise }\end{cases} .
\end{equation}
Note that $u_i(t)$ is quantized to a ternary value. The set $ \mathcal{U}(t)=\left\{i \in \mathcal{V}: \theta_i(t)=0 \right\}$ contains the agents whose local clock variables are equal to zero and the map $f_{\varepsilon}(x)$ is defined as
\begin{equation*}
f_{\varepsilon}(x)= \begin{cases}x & \text { if }|x| \geq \varepsilon \\ 0 & \text { otherwise. }\end{cases}
\end{equation*}
where the sensitivity parameter $\varepsilon$ is a positive scalar. Additionally, the function $\operatorname{ave}_i(t)$ is mathematically expressed as
\begin{equation*}
\operatorname{ave}_i(t)=\sum_{j \in \mathcal{R}_i^+(t)} a_{i j}(t)\left(\hat{x}_j\left(t-\tau_j^i(t)\right)-x_i(t)\right),
\end{equation*}
where $\tau_j^i(t)$ represents the time delay. Define $e_i(t)=t-t_m^i$ as the time interval since agent $i$'s last update. Suppose that there exists an upper limit $\tau^{\prime}$ with $e_i(t)+\tau_j^i(t) \leq \tau^{\prime}$. Then, the main result for the self-triggered case is presented below \cite{matsume2021resilient}.

\begin{The}
\label{resultself}
    Consider a time-invariant graph $\mathcal{G}=(\mathcal{V}, \mathcal{E})$ consisting of $n$ agents. Suppose that the network satisfies the $F$-total malicious model and the normal agents execute the E-MSR algorithm for update. Then, resilient consensus at an error level $c$ is achieved if the underlying network is $(2F +1)$-robust. Furthermore, the error level $c$ is accomplished if the parameter $\varepsilon$ satisfies
    \begin{equation*}
        \varepsilon \leq \frac{\omega^{(\tau^{\prime}+1)n-1} (1-\omega) c}{1-\omega^{(\tau^{\prime}+1)n-1}}.
    \end{equation*}
\end{The}

Compared to the theoretical result of the event-triggered case (Theorem~\ref{resultevent}), Theorem~\ref{resultself} merely provides a sufficient condition, but is more general since it takes time delays into consideration. It is also worth mentioning that the work \cite{senejohnny2019resilience} utilized the self-triggered mechanism to devise an MSR-type protocol, which is resilient against node injection attacks in asynchronous networks. Furthermore, the work \cite{senejohnny2019resilience} presents the condition on network topology in terms of the number of in-neighbors and common in-neighbors. This graph condition is more applicable for verifying network properties in the context of large-scale networks.

\subsection{Quantized Communication Mechanism}
In the context of MASs, quantized communication refers to a mechanism where agents exchange discrete or quantized information rather than continuous values before transmission, which is approximate for limited capabilities in communications and computations of the agents. It is an effective way to reduce the data size to be sent in each transmission and has been widely studied in consensus problems \cite{kashyap2007quantized,nedic2009distributed1,carli2010gossip,lavaei2011quantized,gravelle2014quantized,el2016design}. In this subsection, we focus on the quantized resilient consensus problem, where the MAS is subject to node injection attacks and agents adopt integer-valued states for update \cite{dibaji2017resilient2,wang2020event}.

Consider a quantized MAS where limited capabilities in communications and computations of the agents force them to take integer values. All agents utilize states $x_i[t] \in \mathbb{Z}$ and inputs $u_i[t] \in \mathbb{Z}$ at each time step $t \in \mathbb{Z}_{\geq 0}$. Different from the real-valued case, the quantized version requires the update rule to be randomized, which motivates us to introduce the quantization function $Q : \mathbb{R} \rightarrow \mathbb{Z}$. Specifically, a probabilistic manner is executed as \cite{aysal2008distributed}
\begin{equation}
Q(y)= \begin{cases}\lfloor y\rfloor & \text {with probability } p(y), \\ \lceil y\rceil & \text {with probability } 1-p(y),\end{cases}
\label{Quantizer}
\end{equation}
where $p(y)=\lceil y\rceil-y$, and the floor function  $y$  gives the greatest integer less than or equal to $y$.

With the introduction of (\ref{Quantizer}), the consensus condition in Definition~\ref{Problemerc} should be presented in a probabilistic manner. In particular, the normal agents will try to solve the following resilient quantized consensus problem.
\begin{Def}[\textit{Resilient quantized consensus}]
\label{Problemerqc}
    Consider a graph $\mathcal{G}=(\mathcal{V}, \mathcal{E})$. Suppose that the network satisfies a certain threat model ($F$-total, $F$-local or $f$-fraction local). For any choice of initial values, determine graph conditions and design controllers such that the following conditions are satisfied.
\begin{itemize}
    \item Resilience condition: For each normal agent $i \in \mathcal{N}$ and for all time steps $t \in \mathbb{Z}_{\geq 0}$, it holds $x_i[t] \in \mathcal{S}$.
    \item Consensus condition: For each normal agent $i \in \mathcal{N}$, there exists a finite time step $t_a \geq 0$ such that
    \begin{equation*}
        \operatorname{Prob}\left\{x_i\left[t_a\right] \in \mathcal{C}_{|\mathcal{N}|} \mid x[0]\right\}=1,
    \end{equation*}
    where the consensus set $\mathcal{C}_{|\mathcal{N}|}$ is defined as
    \begin{equation*}
        \mathcal{C}_{|\mathcal{N}|}=\left\{ x \in \mathbb{Z}^{|\mathcal{N}|} \mid x_1=\cdots=x_{|\mathcal{N}|} \right\}.
    \end{equation*}
\end{itemize}
\end{Def}

\begin{table*}[!t]
\caption{Summary of Key References on Resilient Coordination with Different Communication Mechanisms.}
\label{Tab4}
\centering
\resizebox{1\linewidth}{!}{
\begin{tabular}{|c|c|c|c|c|c|c|}
\hline
References         & \begin{tabular}[c]{@{}c@{}}Communication \\ Mechanism\end{tabular}            & Time Domain                    & Consensus Type                         & Resilient Algorithm     & Attack Model & Graph Condition    \\ \hline
\cite{wang2019resilient}                  & \begin{tabular}[c]{@{}c@{}}Event-triggered\\ strategy\end{tabular}            & Discrete-time                  & Approximate consensus                  & E-MSR                   & $F$-total    & $(F+1,F+1)$-robust \\ \hline
\cite{matsume2021resilient}                  & \begin{tabular}[c]{@{}c@{}}Self-triggered\\ strategy\end{tabular}             & Continuous-time                & Exact consensus                        & E-MSR                   & $F$-total    & $(2F+1)$-robust    \\ \hline
\multirow{2}{*}{ \cite{dibaji2017resilient2} } & \multirow{2}{*}{\begin{tabular}[c]{@{}c@{}}Quantized\\ strategy\end{tabular}} & \multirow{2}{*}{Discrete-time} & \multirow{2}{*}{Approximate consensus} & \multirow{2}{*}{QW-MSR} & $F$-total    & $(F+1,F+1)$-robust \\ \cline{6-7} 
                   &                                                                               &                                &                                        &                         & $F$-local    & $(2F+1)$-robust    \\ \hline
\end{tabular}
}
\end{table*}

In the presence of node injection attack, the paper \cite{dibaji2017resilient2} proposed a \textit{Quantized Weighted Mean-Subsequence-Reduced} (QW-MSR) algorithm. The main steps of the QW-MSR algorithm are the same as those of the W-MSR algorithm, except for the update rule. Since the probabilistic quantizer is involved, the quantized control protocol for agent $i$ is mathematically expressed as
\begin{equation}
u_i[t]=Q\left(\sum_{j \in \mathcal{R}_i^+[t]} a_{i j}[t]\left(x_j[t]-x_i[t]\right)\right).
\label{schemequantized}
\end{equation}

It is worth mentioning that the probabilistic quantizer deployed on each agent is independent and may vary from agent to agent. Therefore, the control protocol (\ref{schemequantized}) is able to be executed in a distributed manner. Intuitively, the probabilistic quantization is equivalent to the widely recognized technique of dithering \cite{aysal2008distributed}, which finds extensive applications in the field of signal processing.

With the QW-MSR algorithm, the main result for the quantized case is presented below.
\begin{The}
\label{mainresultquantized}
    Consider a time-invariant graph $\mathcal{G}=(\mathcal{V}, \mathcal{E})$, where each agent takes integer values for update. Suppose that the network satisfies the $F$-total malicious model and the normal agents execute the QW-MSR algorithm for update. Then, resilient quantized consensus is guaranteed if and only if the underlying network is $(F + 1, F + 1)$-robust.
\end{The}

The condition on network topology in Theorem~\ref{mainresultquantized} is identical with Theorems~\ref{mainresultsingle},~\ref{mainresultdouble}, and \ref{mainresultnonlinear} for the real-valued agent cases with single-integrator, double-integrator, and leader-follower structures, respectively. Nevertheless, the quantized case investigates whether agents could achieve convergence within a finite timeframe and in a probabilistic manner. It was further revealed in \cite{dibaji2017resilient2} that a $(2F+1)$-robust graph is a sufficient condition for guaranteeing resilient quantized consensus with the QW-MSR algorithm, when the system is subject to the $F$-local malicious attack.

TABLE~\ref{Tab4} provides a summary of the key references on resilient coordination with different communication mechanisms.

\section{Application Scenarios}
\label{sec6}
In this section, we will review some other resilient cooperative tasks originated from resilient consensus problems and their applications in practical scenarios. Specifically, the notions of \textit{resilient containment} and \textit{resilient distributed optimization} are presented, and we will examine their applications in multi-robot systems and power systems, respectively. Furthermore, some other application-oriented extensions related to resilient coordination are further reviewed with the corresponding literature for reference.

\subsection{Resilient Containment for Multi-Robot Systems}


Containment control, as a method of the multi-agent cooperative control, is an important area of research in the field of CPSs and has attracted remarkable attention in the past decade. Different from achieving consensus, resilient containment problems generally require a leader-follower structure of the MAS and seek for appropriate distributed algorithms to drive the follower agents (followers) to move within a containment area constructed by the leader agents (leaders) despite the influence of node injection attacks.


In the context of resilient containment, the state values of the leaders define a containment area. The followers are expected to reach this area and not escape from it, even in the presence of node injection attacks. Specifically, resilient containment aims to solve the following problem.
\begin{Def}[\textit{Resilient containment}]
Consider a leader-follower MAS described by a graph $\mathcal{G}=(\mathcal{V}, \mathcal{E})$ and a containment area $\mathcal{C}_l$ defined according to the position of the leaders. Suppose that the network satisfies a certain threat model ($F$-total, $F$-local or $f$-fraction local). For any choice of initial values, determine graph conditions and design controllers such that $\lim _{t \rightarrow \infty} x_i[t] \in \mathcal{C}_{l}, \ \forall i \in \mathcal{N}$.
\end{Def}


Note that the containment area $\mathcal{C}_l$ is completely determined by leaders and may be set at an arbitrary position. In addition, the achievement of resilient containment requires $|\mathcal{L}| \geq 2$, so that $\mathcal{C}_l$ can be constructed. In one-dimensional space \cite{yan2020resilient}, the containment area $\mathcal{C}_l$ is a safety interval defined by $\mathcal{S}=\left[\min_{l \in \mathcal{L}_n} x_l[t], \max_{l \in \mathcal{L}_n} x_l[t] \right]$, while in two- and three-dimensional spaces \cite{santilli2021dynamic,santilli2022secure}, $\mathcal{C}_l$ becomes a finite plane and a polyhedron, respectively.

In \cite{yan2020resilient}, resilient containment is guaranteed for single-integrator and double-integrator MASs, where the leaders are assumed to be static. Specifically, the idea of W-MSR is adopted and the control protocols for first-order agents is designed as
\begin{equation}
    u_i[t]=-\sum_{j \in \mathcal{R}_i^+[t]} a_{i j}[t]\left(x_i[t]-x_j[t]\right),\
    \label{first}
\end{equation}
and the protocol for second-order agents is given by
\begin{equation}
u_i[t]=-2 \rho v_i[t]-\sum_{j \in \mathcal{R}_i^+[t]} a_{i j}[t]\left(x_i[t]-x_j[t]\right),
\label{second}
\end{equation}
where $\rho$ is a positive scalar. Note that $\mathcal{R}_i^+[t]$ in (\ref{first}) and (\ref{second}) means that both first-order and second-order agents will only utilize the state values of in-neighbors retained after the W-MSR algorithm for update.

\begin{The}
\label{resultcontainment}
    Consider a single-integrator MAS described by a time-invariant graph $\mathcal{G}=(\mathcal{V}, \mathcal{E})$. Suppose that the network satisfies the $F$-local malicious model and the normal agents execute (\ref{first}) and (\ref{second}) for update. Then, resilient containment is achieved if the underlying network is strongly $(3F+1)$-robust w.r.t. $\mathcal{L}$.
\end{The}

The sufficient condition on network topology for double-integrator case is the same as Theorem~\ref{resultcontainment}, except for an additional parameter condition imposed on $\rho$. Let $\sqrt{1-\omega} < \rho < 1$ hold, where $\omega$ is the lower bound of weight $a_{i j}[t]$. The authors in \cite{yan2020resilient} proved that resilient containment is achieved for the double-integrator MAS if the underlying network is strongly $(3F+1)$-robust w.r.t. $\mathcal{L}$.

\begin{Rem}
    According to Theorem~\ref{resultcontainment}, to achieve resilient containment, the network topology should be strongly $(3F+1)$-robust w.r.t $\mathcal{L}$. However, this property has high requirements on network connectivity and complexity, which may be difficult to satisfy, especially in large-scale distributed networks. Additionally, the paper \cite{usevitch2018resilient} revealed that strong robustness induces a lower bound on the number of leaders, e.g., if the network is strongly $r$-robust w.r.t. $\mathcal{L}$, then $|\mathcal{L}|\geq r$. Thus, how to relax or remove the stringent constraint on network topology and achieve resilient containment with relaxed graph robustness still remain open questions.
\end{Rem}

\begin{Rem}
    By comparing the protocols (\ref{first}) and (\ref{second}) with (\ref{scheme2}) and (\ref{control input double-integrator}), we can find that they are essentially identical expressions but serving for different resilient objectives. Intuitively, if we let all leaders maintain the same static state, the resilient containment problem will become the leader-follower resilient consensus problem, where the normal followers are expected to reach agreement on a consensus value determined by normal leaders.
\end{Rem}

Multi-robot systems (MRSs) are widely recognized as the collective term for MASs that incorporates robotic agents or swarms. One notable advantage of MRSs lies in their cost-effectiveness when deploying a substantial quantity of diverse robotic units. These units are capable of undertaking tasks that would be beyond human capabilities or too perilous for human involvement. Examples include environmental exploration and search and rescue operations in adversarial environments. However, lack of global situational awareness makes distributed MRSs vulnerable to malicious attacks or faults. Thus, in the following part, we focus on the resilient containment control problem for MRSs. The objective is to drive a set of followers to reach and then remain confined within an area defined by the positions of a set of possibly moving leaders, despite the influence of malicious agents. During this process, robots may suffer a cyber attack or encounter a fault, thereby failing to execute the prescribed control protocols and achieve containment.

In \cite{santilli2022secure}, a secure static containment strategy for MRSs with adversarial intruders was proposed. The authors developed a distributed local interaction protocol for MRSs in a multi-dimensional space. The protocol is designed to operate within a time-varying digraph and aims to reach resilient containment within the convex hull of a certain group of leaders. Specifically, the convex analysis is adopted and the notion of \textit{convex hull} is defined as
\begin{equation*}
\operatorname{co}(\mathbf{x}, \mathcal{S})=\left\{y \in \mathbb{R}^d, y=\sum_{i \in \mathcal{S}} \alpha_i x_i: \sum_{i \in \mathcal{S}} \alpha_i=1, \alpha_i \geq 0\right\},
\end{equation*}
where $\operatorname{co}(\mathbf{x}, \mathcal{S})$ is a compact set which contains the state values of the agents belong to a subset $\mathcal{S} \subseteq \mathcal{V}$. 

The \textit{secure convex hull} $\operatorname{\mathrm{s}-co}(\mathbf{x})$ is defined as $\operatorname{\mathrm{s}-co}(\mathbf{x})=\operatorname{co}\left(\mathbf{x}, \mathcal{N}\right)$ and the \textit{$F$-secure convex hull} $\operatorname{\mathrm{F}-co}(\mathbf{x},\mathcal{V}_i^{+})$ is given by
\begin{equation}
    \operatorname{\mathrm{F}-co}(\mathbf{x},\mathcal{V}_i^{+})=\bigcap_{\substack{\forall \mathcal{I} \subseteq \mathcal{V}_i^{+} \\|\mathcal{I}|=\min \left\{F,\left|\mathcal{V}_i^{+}\right|\right\}}} \operatorname{co}\left(\mathbf{x}, \mathcal{V}_i^{+} \cup \{i\} \backslash \mathcal{I}\right), 
    \label{Fsecure}
\end{equation}
where the former represents the convex hull constructed by all normal agents, and the latter is constructed by the intersection of the convex hull of the state values of the in-neighbors for agent $i$, where a different subset of the in-neighbors of cardinality equal to $\min \left\{F,\left|\mathcal{V}_i^{+}\right|\right\}$ is removed each time.

To achieve resilient containment in a multi-dimensional space, each agent implements the following $F$-secure containment algorithm:
\begin{enumerate}
    \item \textit{(Collecting in-neighbors' information):} At each time step $t$, each normal agent $i \in \mathcal{N}$ broadcasts $x_i[t]$ to its out-neighbors, receives $x_j[t]$ from its in-neighbors $j \in \mathcal{V}_i^+[t]$, and collects $x_i[t]-x_j[t]$ in its own reference frame.
    \item \textit{(Computing the $F$-secure convex hull):} Compute the $F$-secure convex hull at time step $t$ according to (\ref{Fsecure}).
    \item \textit{(Computing the centroid):} Compute the centroid $m_i[t]$ of the $F$-secure convex hull at time step $t$ according to
    \begin{equation*}
        m_i[t]=\mathrm{C}\left(\operatorname{\mathrm{F}-co}(\mathbf{x}[t],\mathcal{V}_i^{+}[t])\right),
    \end{equation*}
    where $\mathrm{C}\left(\operatorname{\mathrm{F}-co}(\mathbf{x}[t],\mathcal{V}_i^{+}[t])\right)$ is the centroid of the $F$-secure convex hull computed w.r.t. the local reference frame for agent $i$.
    \item \textit{(Updating local states):} Update the state value for agent $i$ as
    \begin{equation*}
    x_i[t+1]=(1-\varepsilon)x_i[t]+ \varepsilon m_i[t].
    \end{equation*}
\end{enumerate}

The core of the $F$-secure containment algorithm is that all normal agents are driven towards a designated containment area, which ensures that the movement of normal agents remains unaffected by the misbehavior of adversarial agents. The secure containment area is constructed by intersecting a combination of convex hulls formed by the state values of the in-neighbors. Under appropriate network topologies, the agents can move freely in this containment area and pursue a global objective collaboratively without encountering any disruptions. It is worth noting that the intersection encompasses the state of the agent $i$, which is inherently safe as per its definition, whereas the subset $\mathcal{I}$ does not possess this property.

The main result for resilient containment in a multi-dimensional space is stated as follows.
\begin{The}
    Consider a MRS modelled in a multi-dimensional space and described by a time-varying digraph $\mathcal{G}[t]=(\mathcal{V}, \mathcal{E}[t])$. Suppose that the network satisfies the $F$-local malicious model and the normal agents execute the $F$-secure containment algorithm. Then, resilient containment is achieved if the underlying network is strongly $(F (d + 1) + 1)$-robust w.r.t. $\mathcal{L}$. Furthermore, the containment area $\mathcal{C}_l$ is given by $\mathcal{C}_l=\operatorname{co}(\mathrm{x}[0],\mathcal{L})$.
\end{The}

For situations where the leaders are dynamic, the work \cite{santilli2021dynamic} incorporates a sign function in the control input and achieves resilient containment for the normal followers under adversarial environments. Specifically, the control protocol for the normal followers is described by
\begin{equation}
    u_i^f(t)=-\alpha \sum_{j \in \mathcal{V}_i^{+}} \operatorname{\textit{sign}}\left(x_i(t)-x_j(t)\right), \quad i \in \mathcal{F}_n,
    \label{dynamiccontainmentprotocol}
\end{equation}
where $\alpha$ is a positive control gain.

The following theorem provides a sufficient condition for the achievement of dynamic resilient containment (resilient containment where the leaders are dynamic).

\begin{The}
\label{resultdynamiccontainment}
Consider a MRS described by a time-invariant graph $\mathcal{G}=(\mathcal{V},\mathcal{E})$. The node set $\mathcal{V}$ satisfies $\mathcal{V}=\mathcal{L}\cup\mathcal{F}\cup\mathcal{M}$, with $\mathcal{L}$, $\mathcal{F}$, and $\mathcal{M}$ being nonempty and disjoint node sets. Suppose that the control inputs of the leaders are bounded by $u_{\text{max}}^l \in \mathbb{R}_{\geq 0}$ and the normal followers execute (\ref{dynamiccontainmentprotocol}) for update. Then, resilient dynamic containment is achieved after a finite time $T >0$ if the underlying network is $(r,s)$-robust and if the conditions below hold:
\begin{equation}
\text{1)} \  \left(\alpha-u_{\max }^{l}\right)\left(r-|\mathcal{M}|\right) \min \left\{|\mathcal{F}|, |\mathcal{L}|, s\right\}>\alpha\left|\left(\mathcal{F},\mathcal{M}\right)\right| ; \quad
\text{2)} \  \alpha > u_{\max }^{l}; \quad
\text{3)} \  r > |\mathcal{M}|.
\label{dynamic condition}
\end{equation}
where $(\mathcal{F},\mathcal{M})=\{(i, j) \in \mathcal{E} : i \in \mathcal{F}, j \in \mathcal{M}\}$ and $r,s \in \mathbb{Z}_{\geq 0}$. Furthermore, the containment area is given by
\begin{equation*}
\mathcal{C}_l(t)=\left\{y \in \mathbb{R}^d: y_p \in\left[\min_{i \in \mathcal{L}} x_{i,p}(t), \max_{i \in \mathcal{L}} x_{i,p}(t)\right]\right\},
\end{equation*}
where $p \in\{1, \ldots, d\}$.
\end{The}

In contrast to the current studies on resilient coordination that involve 1) removing some suspicious values received from in-neighbors and 2) weighting the retained state values based on specific non-negative scalar, the idea of MSR is not adopted in \cite{santilli2021dynamic}. This is because that the conditions in Theorem~\ref{resultdynamiccontainment} have already guaranteed that the eventual direction that the followers move towards is not influenced by adversary agents. 

\begin{Rem}
    Although Theorem~\ref{resultdynamiccontainment} does not explicitly state the specific attack model that the network needs to satisfy, the third condition in (\ref{dynamic condition}) has posed a constraint on the maximum number of malicious agents, which can be regarded as the $(r-1)$-total malicious model. In addition, the containment area $\mathcal{C}_l(t)$ here is a time-dependent function, which characterizes the dynamic property for leaders.
\end{Rem}

\subsection{Resilient Distributed Optimization for Micro Grid}

Distributed optimization refers to the process of optimizing a global objective or a set of individual objectives in a decentralized manner, where multiple agents work collaboratively to find the optimal solution. In this context, agents can represent a variety of entities such as drones \cite{liu2021distributed}, robots \cite{montijano2014efficient}, or even devices in micro grid \cite{maknouninejad2014realizing}. The investigation of distributed optimization has gained significant attention in the past decade. This advancement offers several advantages including higher scalability, stronger robustness, and higher efficiency in comparison to centralized patterns \cite{yang2019survey}. Compared with the consensus problem, distributed optimization possesses a more general setting, where multiple agents are equipped with local cost functions and aim to optimize a global objective concerning with these functions.

Currently, an extensive amount of work has been dedicated to the investigation and analysis of distributed optimization methods based on consensus approach and subgradient descent technique \cite{nedic2009distributed, nedic2010constrained,shi2015extra,varagnolo2015newton}. Nevertheless, these methods are built on the predominant assumption that all agents seek for the global optimizer cooperatively, while the distributed property makes CPSs vulnerable to external malicious attacks. It was revealed in \cite{sundaram2018distributed} that even a single malicious agent may compel all agents to reach arbitrary non-optimal values by merely keeping this value constant, thus failing to achieve the global optimum. Thus, it is critical to study resilient coordination for CPSs by designing reliable and secure algorithms to achieve the desired distributed optimization in presence of node injection attacks.

Consider a time-invariant graph $\mathcal{G}=(\mathcal{V}, \mathcal{E})$ consisting of $n$ agents, where each agent $i \in \mathcal{V}$ possesses a local and confidential cost function $f_i(x)$ and its objective is to optimize an average of these cost functions. Thus, all agents will endeavor to tackle the following optimization problem cooperatively:
\begin{equation*}
\mathop{\arg\min}\limits_x F(x) =\min \frac{1}{n} \sum_{i=1}^n f_i(x), \ i \in \mathcal{V}.
\end{equation*}

However, in the presence of node injection attack, the work \cite{sundaram2018distributed} proved that for any distributed optimization algorithms, the optimization result may be manipulated by attackers arbitrarily. Specifically, a single malicious agent is able to affect the eventual result of the distributed optimization while avoiding detection. This fact indicates a tradeoff between resilience and optimality. In other words, a distributed optimization algorithm that finds a global optimizer without node injection attacks may also be compromised by an adversary agent. Based on this fundamental limitation, we present the \textit{resilient distributed optimization} problem below, where the CPS aims to solve a compromised distributed optimization problem.


\begin{Def}[\textit{Resilient distributed optimization}]
   Consider a graph $\mathcal{G}=(\mathcal{V}, \mathcal{E})$ consisting of $n$ agents. Suppose that the network satisfies a certain threat model ($F$-total, $F$-local or $f$-fraction local). For any choice of initial values, determine graph conditions and design controllers such that $\lim _{t \rightarrow \infty} x_i[t] = x^*, \ \forall i \in \mathcal{N}$, where $x^*$ is the global optimal value for the distributed optimization problem
\begin{equation*}
    \mathop{\arg\min}\limits_x F(x) =\min \frac{1}{|\mathcal{N}|} \sum_{i \in \mathcal{N}} f_i(x).
\end{equation*}
\end{Def}

In \cite{sundaram2018distributed}, the authors made a mild assumption that each local cost function $f_i(x), \ \forall i \in \mathcal{V}$ is locally Lipschitz and convex with bounded subgradients. In the presence of $F$-total malicious model, a consensus-based resilient distributed optimization protocol is developed as
\begin{equation}
    u_i[t]=-\sum_{j \in \mathcal{R}_i^+[t]} a_{i j}[t]\left(x_i[t]-x_j[t]\right)-\alpha_t d_i[t],
    \label{distributedoptimizationprotocol}
\end{equation}
where $d_i[t]$ is a subgradient of $f_i(x)$ and $\alpha_t$ is a diminishing step size.


Note that protocol (\ref{distributedoptimizationprotocol}) consists of the consensus term $-\sum_{j \in \mathcal{R}_i^+[t]} a_{i j}[t]\left(x_i[t]-x_j[t]\right)$ and the subgradient term $-\alpha_t d_i[t]$, while $\mathcal{R}_i^+[t]$ means that the local filtering (LF) operation \cite{sundaram2016secure} is adopted to remove the suspicious values received from in-neighbors. Thus, this approach incorporates the consensus-based and gradient tracking techniques, and utilizes the idea of MSR to defend against the malicious attack.

The work \cite{fu2021resilient} provides the necessary and sufficient conditions to guarantee convergence for the CPS under adversarial environments.

\begin{The}
\label{Theo1}
Consider a time-invariant graph $\mathcal{G}=(\mathcal{V}, \mathcal{E})$. Suppose that the network satisfies the $F$-local malicious model and the normal agents execute (\ref{distributedoptimizationprotocol}) for update. Let $\lim_{t \rightarrow \infty}\alpha_t=0$. Then, the final states of all normal agents converge to a consensus value if and only if the underlying network is $(2F+1)$-robust.
\end{The}

Note that Theorem~\ref{Theo1} merely guarantees that the CPS reaches an agreement under the $F$-local malicious model. Suppose that each local cost function $f_i(x)$ possesses a nonempty compact set of minimizers $\mathcal{X}^{*}_i$. Furthermore, let $m^{*}=\min_{i \in \mathcal{N}}\min \{ x| x \in \mathcal{X}^{*}_i \}$ and $M^{*}=\max_{i \in \mathcal{N}}\max \{ x| x \in \mathcal{X}^{*}_i \}$. The following theorem reveals that, despite malicious attacks, the state values of normal agents will converge to an optimal interval $\Psi = [m^*,M^*]$, which refers to the convex hull of local minimizers. Resilient distributed optimization is thereby guaranteed.
\begin{The}
    Consider a time-invariant graph $\mathcal{G}=(\mathcal{V}, \mathcal{E})$.  Suppose that the network satisfies the $F$-local malicious model and the normal agents execute (\ref{distributedoptimizationprotocol}) for state update. Let $\lim_{t \rightarrow \infty}\alpha_t=0$ and $\sum_{t=0}^{\infty} \alpha_t=\infty$. Then, the consensus value stated in Theorem~\ref{Theo1} will lie in the optimal interval $\Psi = [m^*,M^*]$ if the underlying network is $(2F+1)$-robust.
    \label{Theo4}
\end{The}

\begin{Rem}
    Although Theorems~\ref{Theo1} and \ref{Theo4} guarantee the resilience and optimality for CPSs, respectively, they do not attain accurate distributed optimization under two attack scenarios. Rather than seeking for convergence to a safety interval, one may ask whether it is achievable that the CPS exactly converges to the global optimizer under adversarial environments. The paper \cite{sundaram2018distributed} claimed that this goal is unattainable unless extra assumptions are satisfied. Motivated by this result, the paper \cite{gupta2021byzantine} showed that accurate distributed optimization is achieved under adversarial environments if the $2F$-redundancy assumption was fulfilled by local cost functions. Nevertheless, the work \cite{gupta2021byzantine} executed the analysis in a peer-to-peer network, while the attack scenario was limited to the $F$-total model. It was further revealed in \cite{gupta2020fault} that the  assumption is a necessary condition for ensuring accurate optimization. Extending the results \cite{gupta2021byzantine,gupta2020fault} to general networks and providing more complete optimality conditions remain open directions of research.
\end{Rem}

In the context of distributed optimization for micro grids, resilience is also an essential property. The paper \cite{duan2016resilient} explores the vulnerability of the distributed DC optimal power flow (DC-OPF) algorithm to data integrity attacks and presents a secure version of the algorithm with an embedded attack-resilient coordination mechanism. To increase the resilience of a distributed electric system against natural disasters, the work \cite{yuan2016robust} formulates a resilient distribution network planning problem (RDNP) to coordinate the hardening and distributed generation (DG) resource placement with the objective of minimizing the system damage. In \cite{gallo2018distributed}, a distributed observer structure is constructed to detect the potential FDI risks in microgrids. Subsequently, the malicious data channels that have been detected will be isolated. Other relevant work can be found in \cite{hao2015sparse,singh2017joint,ding2019distributed}. Nevertheless, most of these promising studies focus on detecting the infected components in the smart grid and subsequently isolating these threats, while it is impossible to enumerate or eliminate every potential attack threats for a perfectly secured smart grid. Thus, the concept of attack-resilience should be integrated, and resilient distributed optimization approaches represented by MSR-type algorithms should be developed to defend against the permanent presence and evolution of threats.



\subsection{Other Application-Oriented Extensions Related to Resilient Coordination}

In the past few years, miscellaneous application-oriented extensions have been developed, and various resilient distributed algorithms have been designed to achieve resilient coordination in CPSs. Typical studies related to resilient coordination are reviewed in TABLE~\ref{Tab5}. Although most of them focus on resilient consensus problems, they can be easily extended to other resilient coordination problems by changing system structures or deploying additional components.

\begin{table*}[!t]
\caption{List of Application-Oriented Extensions Related to Resilient Coordination of CPSs.}
\label{Tab5}
\centering
\resizebox{1\linewidth}{!}{
\begin{tabular}{|c|c|c|}
\hline
References & Extension                                                 & Key Idea                                                                                                                                                                          \\ \hline
\cite{shang2023resilientmulti,abbas2022resilient,yan2022resilient}          & Resilient vector consensus                                &  \begin{tabular}[c]{@{}c@{}}Design scalable resilient algorithms to solve resilient coordination problems \\ in a multi-dimensional space.\end{tabular}                                                                                                                                                    \\ \hline
\cite{liu2018resilient,wu2021resilient}          & Resilient bipartite consensus                             & \begin{tabular}[c]{@{}c@{}}Model a signed network and design secure controllers \\ such that the normal agents converge to two opposite values.\end{tabular}                                                                                            \\ \hline
\cite{fiore2019resilient,ying2023privacy,gusrialdi2023resilient}          & Resilient privacy-preserving consensus                    &  \begin{tabular}[c]{@{}c@{}}Develop resilient algorithms to satisfy differential privacy requirement \\ and guarantee resilient consensus for normal agents.\end{tabular}                                                                                                                                 \\ \hline
\cite{abbas2014resilient,mitra2018impact}         & \begin{tabular}[c]{@{}c@{}}Resilient consensus based on \\ trusted nodes\end{tabular}                & \begin{tabular}[c]{@{}c@{}}Introduce trusted nodes and design resilient algorithms to \\ defend against any number of malicious agents.\end{tabular}                                                                                                                                                        \\ \hline
\cite{shang2021median,shang2023resilient}          & \begin{tabular}[c]{@{}c@{}}Resilient consensus over \\ random networks\end{tabular}                  & \begin{tabular}[c]{@{}c@{}}Model a time-varying random digraph and proposed a \\ median-based strategy for ensuring consensus.\end{tabular}                                                                                                                                                                               \\ \hline
\cite{d2016resilient,yuan2023event}          & \begin{tabular}[c]{@{}c@{}}Resilient consensus with \\ multi-hop communication\end{tabular}          & \begin{tabular}[c]{@{}c@{}}Adopt a multi-hop communication mechanism and design two  event-based \\ controllers to solve the approximate resilient consensus problem.\end{tabular}                                                                                                                                                                                  \\ \hline
\cite{wang2022resilientmobile,wang2021resilientmobile2}          & \begin{tabular}[c]{@{}c@{}}Resilient consensus under \\ adversarial spreading processes\end{tabular} & 
\begin{tabular}[c]{@{}c@{}}Address resilient consensus problems for MASs when the \\ network is subject to spreading malicious attacks.\end{tabular}                                                                                                                                                                                 \\ \hline
\cite{saulnier2017resilient,amirian2021distributed}         & Resilient flocking                                        & \begin{tabular}[c]{@{}c@{}}Devise resilient algorithms that ensure resilient consensus \\ for a group of robots on the direction of movement.\end{tabular}                                                                                                                                                                                 \\ \hline
\cite{li2023resilient,gong2023resilient}          & Resilient formation tracking                              &  \begin{tabular}[c]{@{}c@{}}Investigate time-varying output formation tracking problems for \\ heterogeneous MASs in an adversarial environment.\end{tabular}                                                                                                                                                                                 \\ \hline
\cite{yan2021resilient,peng2023resilient}         & Resilient output regulation                               & \begin{tabular}[c]{@{}c@{}}Design resilient algorithms such that the normal agents asymptotically \\ track preset trajectories or reject disturbances.\end{tabular} \\ \hline
\cite{chen2018resilient,an2021byzantine}         & Resilient distributed estimation                          &   \begin{tabular}[c]{@{}c@{}}Estimate/track undisclosed states obtained from a network of sensors \\ that may operate in an adversarial environment.\end{tabular}                                                                                                                                                                                \\ \hline
\end{tabular}
}
\end{table*}

\section{Conclusion and Future Directions}
\label{sec7}
This survey provides an in-depth overview of recent achievements of resilient coordination for CPSs. To begin with, we partition the CPS into a hierarchical framework: physical layer, network layer, and communication layer. Based on this hierarchical framework, we summarize the existing work on resilient coordination from three perspectives: physical structure, network topology, and communication mechanism. From the perspective of physical structure, we review several state-of-the-art works aimed at achieving resilient consensus under different system models. In view of the network topology, we mainly focus on resilient consensus problems over time-varying networks and examine three different time-varying versions of robust graphs. Furthermore, the relationships between these three graphs are further discussed. For the communication layer, we introduce three typical communication mechanisms: event-triggered, self-triggered, and quantized mechanisms, all of which are designed to mitigate heavy communication overheads of the system. Therefore, resilient coordination is able to be achieved with low communication burden.

In addition to resilient consensus, we review two other application-oriented resilient coordination tasks: resilient containment and resilient distributed optimization. For resilient containment, we consider two cases that involve multi-dimensional space and dynamic leaders, respectively. These cases apply to multi-robot systems and guarantee that the normal followers converge within a safe area constructed by the leaders, regardless of the influence of node injection attacks. For resilient distributed optimization, based on a fundamental limitation, we introduce a consensus-based optimization method, which incorporates the gradient tracking technique and utilizes the idea of MSR to defend against the malicious attack. Moreover, we discuss the widespread application of distributed optimization technology in micro grids and reveal the importance of implementing resilient distributed optimization in smart grids. Some other application-oriented extensions related to resilient coordination, including resilient formation, resilient flocking, resilient distributed estimation, etc., are further reviewed briefly.

In conclusion, it is apparent that the significance of CPS security will persistently escalate, particularly in light of the escalating frequency of malicious attack occurrences in various applications. Moreover, it is worth mentioning that existing resilient coordination approaches show fundamental limitations because of stringent assumptions and specific requirements on graph robustness. In the context of resilient coordination for CPSs, we want to highlight the following potential avenues for future research:
\begin{itemize}
    \item \textit{Resilient coordination with nonlinear dynamics:} In Section~\ref{sec3}, we discuss resilient consensus problems with different system structures. Nevertheless, all of them involve linear system dynamics. Compared to linear MASs, nonlinear systems capture more complex and realistic dynamics, making them suitable for modeling real-world phenomena that exhibit nonlinear behaviors. Although the work \cite{oksuz2019resilient} incorporated a nonlinear function into the controller, some other nonlinear factors, such as heterogeneity, time delays, and stochasticity should be considered comprehensively. Therefore, an interesting future research direction in resilient coordination is to extend the idea of MSR to nonlinear systems and take more nonlinear analysis tools into account.
    \item \textit{Resilient coordination in a multi-dimensional space:} Expanding existing results to a multi-dimensional space is another trend in resilient coordination. Multi-dimensional resilient consensus problems have been recently studied in \cite{wang2019resilientmulti,yan2020safe,luo2023secure}. In Section~\ref{sec6}, we also discussed resilient containment problems in a multi-dimensional space. For resilient distributed optimization, the multidimensionality still plays an essential role since it not only allows for a more comprehensive exploration of solution spaces, but also provides the possibility to solve the multi-objective optimization problems under adversarial environments.
    \item \textit{Resilient coordination with formal methods:} Resilient coordination often involve specifying complex requirements and spatio-temporal constraints. Formal method, including the linear temporal logic (LTL) \cite{haesaert2020robust}, provides a formal language for precisely expressing these requirements, ensuring that they are unambiguous and clearly defined. Furthermore, resilient coordination strategies need to be verified to ensure they meet the desired safety and performance goals. Formal methods allow for the mathematical verification of control strategies, providing a rigorous way to validate that CPS behaves as intended and adhere to specified spatio-temporal requirements. Therefore, the interplay between resilient coordination and formal method would be of significant interest.
\end{itemize}


\bibliographystyle{ACM-Reference-Format}
\bibliography{sample-base}


\end{document}